\documentclass[a4paper, 12pt]{article}
\usepackage{a4}
\usepackage{amsfonts}
\usepackage{amssymb}
\usepackage{cite}
\usepackage{epsfig}

\author{}

\newcommand{\be}{\begin{equation}}
\newcommand{\ee}{\end{equation}}
\newcommand{\bea}{\begin{eqnarray}}
\newcommand{\eea}{\end{eqnarray}}

\newcommand{\half}{\frac{1}{2}}
\newcommand{\1}{{\bf 1}}
\newcommand{\3}{{\bf 3}}
\newcommand{\2}{{\bf 2}}
\newcommand{\4}{{\bf 4}}
\newcommand{\5}{{\bf 5}}
\newcommand{\6}{{\bf 6}}
\newcommand{\7}{{\bf 7}}
\newcommand{\8}{{\bf 8}}
\newcommand{\9}{{\bf 9}}

\newcommand{\Z}{\mathbb{Z}}

\newcommand{\ov}{\overline}

\newcommand{\tr}{{\rm tr}}
\newcommand{\Tr}{{\rm Tr}}
\newcommand{\Anti}{{\bf Anti}}
\newcommand{\Sym}{{\bf Sym}}
\newcommand{\Adj}{{\bf Adj}}
\newcommand{\n}{{\bf n}}
\newcommand{\N}{{\bf N}}
\newcommand{\M}{{\bf M}}

\newcommand{\ch}{{\rm ch}}

\def\IR{\relax{\rm I\kern-.18em R}}

\def\IP{\relax{\rm I\kern-.18em P}}
\def\inbar{\vrule height1.5ex width.4pt depth0pt}
\def\IC{\relax\,\hbox{$\inbar\kern-.3em{\rm C}$}}

\def\K3{{\bf K3}}

\def\ov{\overline}

\begin{document}

\title{
\begin{flushright} \vspace{-2cm}
{\small CERN-PH-TH/2006-249 \\
\vspace{0.1cm}
\small HD-THEP-06-31 \\
\vspace{-0.35cm}
hep-th/0612030} \end{flushright}
\vspace{4.0cm}
Merging
Heterotic Orbifolds and K3 Compactifications with
Line Bundles
}
\vspace{1.5cm}

\author{\small  Gabriele Honecker$^{\heartsuit}$ and
Michele Trapletti$^{\spadesuit}$ }

\date{}

\maketitle
\begin{center}
\emph{$^{\heartsuit}$ PH-TH Division, CERN \\
  CH - 1211 Geneva 23, Switzerland } \\
\vspace{0.2cm}
{\tt Gabriele.Honecker@cern.ch}\\
\vspace{1.0cm}
\emph{$^{\spadesuit}$ Institut f\"ur Theoretische Physik\\
Universit\"at Heidelberg,
Philosophenweg 16 and 19\\
D - 69120 Heidelberg, Germany }\\
\vspace{0.2cm}
{\tt m.trapletti@thphys.uni-heidelberg.de}
\end{center}
\vspace{1.0cm}

\begin{abstract}
\noindent  
We clarify the relation between six-dimensional Abelian orbifold
compactifications of the heterotic string and smooth heterotic $K3$
compactifications with line bundles for both $SO(32)$ and 
$E_8 \times E_8$ gauge groups.
The $T^4/\Z_N$ cases for $N=2,3,4$ are treated exhaustively, and
for $N=6$ some examples are given.
While all $T^4/\Z_2$ and nearly all $T^4/\Z_3$ models have a simple
smooth match involving one line bundle only, this is only true for
some $T^4/\Z_4$ and $T^4/\Z_6$ cases.
We comment on possible matchings with more than one line bundle for 
the remaining cases.\\
The matching is provided by comparisons of the massless spectra and 
their anomalies as well as a field theoretic analysis of the blow-ups.
\end{abstract}

\thispagestyle{empty}
\clearpage

\tableofcontents


\section{Introduction}
\label{Sec:Intro}

Orbifolds of the heterotic string have been known for more than twenty
years~\cite{Dixon:1985jw}, and four dimensional Standard Model
building attempts in this corner of the M-theory star have started to evolve
soon after its implementation~\cite{Ibanez:1987sn}
(see \cite{Buchmuller:2005jr} for recent results).
New interest in the field was then generated by the introduction of the
so-called ``orbifold-GUT'' idea in extensions of the Standard Model
including the presence of extra dimensions~\cite{Kawamura:2000ev}.
Indeed, many of the latest model building attempts were devoted to a
string embedding of this kind of field theory models~\cite{Kobayashi:2004ya}.
Moreover, in~\cite{Lebedev:2006kn} it was shown that the $\Z_6'$ orbifold
limit of $E_8 \times E_8$ is a particularly fertile background to
implement MSSM spectra and compute the supersymmetry breaking with a MSSM
spectrum's frequency of $10^{-2}$ compared to $10^{-9}$ obtained in a
$\Z_2 \times \Z_2$ type IIA orientifold background with intersecting
branes~\cite{Gmeiner:2005vz}.

On the other hand, $E_8 \times E_8$ heterotic compactifications on Calabi-Yau
manifolds with $SU(n)$ background gauge bundles and GUT spectra have become
popular in the last ten years, see e.g.~\cite{Donagi:1998xe} for some early
works, and recent constructions of potentially phenomenologically interesting
$SU(5)$ and $SO(10)$ GUT spectra with Wilson line breaking to the standard
model can be found e.g. in~\cite{Bouchard:2005ag} and~\cite{Braun:2005nv},
respectively.
 
The two avenues to string phenomenology involving singular and smooth heterotic
backgrounds have been pursued essentially independently without making any
obvious connection.
 
Contrariwise, in type II compactifications it is known that
D-branes in an orbifold background can be treated on equal footing as
those in Calabi-Yau backgrounds in the large volume (geometric) regime
by explicitly constructing the cycles at the orbifold point, see
e.g.~\cite{Blumenhagen:2002wn} for the treatment within Intersecting
Brane Worlds, for more references see also~\cite{Blumenhagen:2006ci}.

Furthermore, 
in~\cite{Blumenhagen:2005pm} it was shown that compactifications with
D-branes are S-dual to $SO(32)$ heterotic compactifications with
generic $U(n)$ backgrounds. Using $U(n)$ bundles embedded in $E_8 \times E_8$, 
a large class of standard model and flipped $SU(5)$ like chiral spectra were 
constructed subsequently~\cite{Blumenhagen:2006ux}.\footnote{For more four
dimensional heterotic string compactifications with generic (non)-Abelian see
also~\cite{Distler:1987ee}.}

In this article, we aim at closing the gap between the different model building approaches 
by matching heterotic
$SO(32)$ and $E_8 \times E_8$ orbifold spectra with smooth counter
parts which have $U(1)$ gauge backgrounds. For concreteness, we focus
on perturbative supersymmetric Abelian $T^4/\Z_N$ orbifolds and $K3$
compactifications, both without Wilson lines. The $K3$ cases were treated in full generality
in~\cite{Honecker:2006dt} including H5-branes. Taking into account H5-branes
is also straightforward on the orbifold side as discussed in~\cite{Aldazabal:1997wi}.

This article is organized as follows: 

In section~\ref{Sec:Orbifolds} we present the framework to compute $T^4/\Z_N$
heterotic orbifold vacua. We reobtain the classification of $\Z_2$ and $\Z_3$
models given in~\cite{Aldazabal:1997wi}, and completely classify the $\Z_4$ models,
both in the $SO(32)$ and in the $E_8\times E_8$ case.
For each orbifold we give the complete list of inequivalent shift
embeddings\footnote{In the $E_8\times E_8$ $\Z_4$ case such a result 
was already given in~\cite{Stieberger:1998yi}.}, and the massless spectra of the
inequivalent models, making large use of the classification strategy described
in~\cite{Choi:2004wn}. We also show some examples of $\Z_6$ models.
For each model, we compute the field theoretic anomaly polynomial in
six dimensions in section~\ref{Sec:Ano} and verify its $4 \times 4$ factorization.

In section~\ref{Subsec:K3general}, we present the model building rules
of heterotic $K3$ compactifications with arbitrary gauge backgrounds 
and compare them with the orbifold construction of
section~\ref{Sec:Orbifolds}. We proceed with the analysis of smooth 
$SO(32)$ heterotic string compactifications in
section~\ref{Subsec:SO32bundles}, where we find an explicit match with
the orbifold models via the coefficients in the anomaly polynomial. 
This leads to the idea that a $\Z_N$ orbifold shift vector can be directly 
translated into the smooth embedding of a line bundle $L$ via the 
relation
\bea
\frac{1}{N}(1,\ldots,1,n,0,\ldots,0) \rightarrow
(L, \ldots, L, L^n, 0, \ldots,0),
\label{Eq:Id_Shifts}
\eea
and the second Chern character (instanton number) of $L$ is computed
from the coefficients of the anomaly polynomial at the orbifold point.
In section~\ref{Subsec:SO_Explicit_Matches}, this recipe is
demonstrated by explicitly matching several $SO(32)$ orbifold spectra.
We apply the same line of reasoning to the $E_8 \times E_8$
compactifications in section~\ref{Subsec:E8_K3_U1s} and show some
explicit matches of $E_8 \times E_8$ orbifold spectra in
section~\ref{Subsec:E8matches}.

Up to this point, only the second Chern characters of the line bundles
have been  determined. In section~\ref{Sec:Expl_Line} we speculate on the 
detailed form of such line bundles.

The ansatz for the orbifold / smooth matching is justified further in
section~\ref{Sec:Blowup} via a field theoretical treatment of the
blow-up procedure. 
We explicitly compute the flat directions of the twisted matter
potential and show that the blow-up procedure, i.e. switching
on a vev along some flat direction, takes care of some seeming
mismatch between orbifold and smooth models.

Finally, in section~\ref{Sec:Conclusions} we conclude, and in
appendix~\ref{App:Decomp} we collect some technical details on gauge symmetry breaking in our models.


\section{Perturbative $T^4/\Z_N$ orbifolds of the heterotic string}
\label{Sec:Orbifolds}

A perturbative $T^4/\Z_N$ heterotic orbifold \cite{Dixon:1985jw}
(see also \cite{Ibanez:1987pj} for the algorithm), in absence
of Wilson lines, is completely specified by the action of the
orbifold operator in the geometric space and in the gauge bundle.
The space-time part of any $\Z_N$ Abelian orbifold action is given by
\bea
\theta: z^i \rightarrow e^{2\pi i v_i} z^i 
\eea
with the shift vector
\bea
{\bf v} \equiv(v_1,v_2,\ldots) =\frac{1}{N}(\sigma_1,\sigma_2,\ldots),
\eea
where $\sigma_i$ are integers subject to the supersymmetry constraint
\bea
\sum_i \sigma_i = 0 \,\, \mbox{mod} \,\,  2.
\eea
In this article, we focus on the supersymmetric $T^4/\Z_N$ orbifolds
with space-time shift vectors
\bea
{\bf v} =\frac{1}{N}(1,-1), \quad\quad N=2,3,4,6.
\eea
The Abelian orbifold action is embedded into the gauge degrees of
freedom by another shift vector
\bea
{\bf V}\equiv(V_1,\ldots,V_{16}) =
  \frac{1}{N}(\Sigma_1, \ldots, \Sigma_{16}),
\eea
with $\Sigma_i$ integer numbers. These are the so called ``vectorial 
shifts''. The embedding is such that a state with weight vector
${\bf w}$ transforms under the orbifold action with a phase given by
$e^{2\pi i{\bf V}\cdot {\bf w}}$. 
In the even order orbifolds of $SO(32)$ string we also take into
account ``spinorial shifts'', i.e. vectors of the form
\bea
{\bf V_S}
= \frac{1}{2N}(\Sigma_1, \ldots, \Sigma_{16})
\eea
with $\Sigma_i$ odd integers.

Since the rotation involves spinors also in the gauge
bundle, the orbifold action has order $N$ only provided that
\bea
N \, \sum_i V_i = 0 \,\, \mbox{mod} \,\, 2.
\label{Eq:Orbi_Modular}
\eea
Moreover, modular invariance of the partition function,
or, equivalently,  the level matching condition that
combines space-time and gauge shifts, requires
\bea
N \left( \sum_i V_i^2 -  \sum_i v_i^2 \right)= 0 \,\, \mbox{mod} \,\, 2.
\label{Eq:Orbi_Level}
\eea
Given these conditions we can classify the possible supersymmetric
$T^4/\Z_N$ heterotic orbifolds and  compute the spectrum of each
model.

\subsection{$SO(32)$ models}
\label{Sec:SO}

\subsubsection{Gauge symmetry breaking and untwisted spectrum}
In a $\Z_N$ orbifold the gauge bosons that are left invariant by the
orbifold action are those with weight vector ${\bf w}$ such that
${\bf V}\cdot {\bf w} = 0 \; {\rm mod} \; 1$.
In the $SO(32)$ case, the weight vectors of the gauge bosons contain only
two non-zero entries which have values $\pm 1$,
$(\underline{\pm 1, \pm 1, 0^{14}})$. Thus, in the case of a
``vectorial'' shift vector \mbox{${\bf V} = (\Sigma_1,\dots,\Sigma_{16})/N$},
an entry $\Sigma_i$ is equivalent, for what concerns gauge symmetry
breaking, to an entry $\Sigma_i+N$ or $-\Sigma_i$.\footnote{Notice
that the replacement $\Sigma_i\rightarrow \Sigma_i+N$ in the shift
vector of a model, harmless for what concerns its untwisted spectrum, 
may modify its twisted matter content, and map it into a different
(inequivalent) model, as we comment later.}
In case $N$ is even, a generic shift vector has the form\footnote{The
superscripts denote the number of identical entries in the shift
vector.} 
\bea
{\bf V}=\frac{1}{N}
\left(0^{n_0},1^{n_1},\dots,\frac{N}{2}^{n_{N/2}}\right)
\eea
and produces the gauge symmetry breaking
\bea
SO(32)\rightarrow SO(2n_0)\times U(n_1)\times\dots
\times U(n_{N/2-1})\times SO(2n_{N/2}).
\eea
The untwisted six dimensional spectrum contains a hyper multiplet in
each of the following representations ($N\neq 2$ cases)\footnote{Multiplets
in six dimensions are CPT invariant, i.e. a hyper multiplet labeled by
${\bf R}$ contains a complex scalar in the representation ${\bf R}$ 
as well as a complex scalar in the conjugate representation $\ov{\bf R}$.}
\bea
&&({\bf 2n_0},{\bf n_1},{\bf 1},\dots {\bf 1})_{1,0,\dots,0},
\nonumber\\
&&({\bf 1},\dots {\bf 1},{\bf n_{N/2-1}},
{\bf 2 n_{N/2}})_{0,\dots,0,1},
\\
&&({\bf 1},\dots,{\bf n_p},\dots,{\bf n_q},\dots,
{\bf 1})_{\dots,1,\dots,1,\dots}\,\,
{\rm for}\,\,p+q=N-1,\,\,q>p,
\nonumber\\
&&({\bf 1},\dots,{\bf n_p},{\bf \ov n_{p+1}},\dots
,{\bf 1})_{\dots,1,-1,\dots}.\nonumber
\eea
The $U(1)$ charges are always zero but for the $U(n)$ with ${\bf n}$
or $\ov{\bf n}$ representations for which it is $+1$ and $-1$,
respectively.
In the $N=2$ case, the gauge symmetry breaking is to
$SO(2n_0)\times SO(2n_1)$, with untwisted matter
$({\bf 2n_0},{\bf 2n_1})$.

In case $N$ is odd, instead, shift vector, gauge symmetry breaking and untwisted
matter are given by
\bea
&&{\bf V}=\frac{1}{N}\left(0^{n_0},1^{n_1},\dots,
        \frac{(N-1)}{2}^{n_{(N-1)/2}}\right),\nonumber\\
&&\nonumber\\
&&SO(32)\rightarrow SO(2n_0)\times U(n_1)\times\dots
        \times U(n_{(N-1)/2}),\nonumber\\
&&({\bf 2n_0},{\bf n_1},{\bf 1},\dots {\bf 1})_{1,0,\dots,0},\nonumber\\
&&({\bf 1},\dots,{\bf n_p},\dots,{\bf n_q},\dots,
        {\bf 1})_{\dots,1,\dots,1,\dots}\,\,
        {\rm for}\,\,p+q=N- 1,\,\,q>p,\\
&&\left({\bf 1},\dots,\frac{\bf n_p (n_p-1)}{\bf 2},\dots,
        {\bf 1}\right)_{\dots,2,\dots}\,\,
        {\rm for}\,\,p=\frac{N-1}{2},\nonumber\\
&&({\bf 1},\dots,{\bf n_p},{\bf \ov n_{p+1}},\dots,
        {\bf 1})_{\dots,1,-1,\dots}.\nonumber
\eea

For ``spinorial'' shift vectors, meaningful only in the even $N$ case,
the vector, for what concerns the low energy gauge group, can always be
brought to the form
\bea
&&{\bf V}_{\bf S}=\frac{1}{2N}\left(1^{n_1},3^{n_3},\dots,(N-1)^{n_{N-1}}\right)
\eea
with gauge symmetry breaking
\bea
SO(32)\rightarrow U(n_1)\times \cdots \times  U(n_{N-1})
\eea
and untwisted matter
\bea
&&\left(\frac{\bf n_1 (n_1-1)}{\bf 2},\dots,{\bf 1}\right)_{2,0\dots,0} ,
\nonumber\\
&&
\left({\bf 1},\dots,\frac{\bf n_{N-1} (n_{N-1}-1)}{\bf 2}\right)_{0\dots,0,2} ,
\nonumber\\
&&({\bf 1},\dots,{\bf n_p},{\bf 1},{\bf \ov n_{p+2}},\dots,
{\bf 1})_{\dots,1,0,-1,\dots}.\nonumber
\eea

\subsubsection{Twisted matter}
In a $\Z_N$ orbifold, extra sectors of {\it twisted} matter are
expected.
In the $\Z_2$ and $\Z_3$ cases we expect the presence of a single
twisted sector in each of the orbifold fixed points.
In the $\Z_4$ case, instead, we expect two twisted sectors. Given $g$
the generator of $T^4/\Z_4$, there is a $g$-twisted spectrum in each
of the four fixed points of $g$ and a $g^2$-twisted spectrum in each
of the 16 fixed points of $g^2$. Similarly, in the $T^4/\Z_6$ case we
expect three twisted sectors: a $g$-twisted one in the single fixed
point of $g$, then a $g^2$-twisted sector in the nine fixed points of
$g^2$ and a  $g^3$-twisted sector in the 16 fixed points of $g^3$.

We do not give the details of the computation of the twisted spectrum
for the $T^4/\Z_2$ and $T^4/\Z_3$ models that have been classified 
in \cite{Aldazabal:1997wi}.
In the $T^4/\Z_4$ case, instead, we discuss in detail the classification
of the models and summarize the twisted spectrum by using the
notation and strategy introduced in~\cite{Choi:2004wn}.

\begin{table}[t]\footnotesize
\renewcommand{\arraystretch}{1.2}
\begin{center}
\begin{tabular}{|c||c||c|c|}
\hline
\hline
$\!\!$\#$\!\!$ & Gauge Group & Untwisted & Twisted \\
& Shift Vector & Matter & Matter
\\\hline\hline
2a & $SO(28) \times SU(2)^2$ & 
$({\bf 28},\2,\2)+4(\1,\1,\1)$ & 
$8({\bf 28},\1,\2)
+ 32(\1,\2,\1)$
\\
& $\half(1^2,0^{14})$ & &
\\\hline
2b & $SO(20) \times SO(12)$ & 
$({\bf 20},{\bf 12})+4(\1,\1)$ & 
$8(\1,{\bf 32_+})$
\\
& $\half(1^6,0^{10})$ & &
\\\hline
2c & $SU(16) \times U(1)$ & 
$2({\bf 120})_2+4(\1)_0$ & 
$16({\bf 16})_{-3}$
\\
& $\frac{1}{4}(1^{15},-3)$ & & 
\\\hline\hline
3a& $SO(28) \times SU(2) \times U(1)$ & 
$({\bf 28},\2)_1+2(\1,\1)_0+1(\1,\1)_2$ & 
$9({\bf 28},\2)_{-1/3}+45(\1,\1)_{2/3}$ 
\\
& $\frac{1}{3}(1^2,0^{14})$ & &  $+18(\1,\1)_{4/3}$
\\\hline
3b & $SO(22) \times SU(5) \times U(1)$ & 
$({\bf 22},\5)_1+(\1,{\bf 10})_2+2(\1,\1)_0$ 
& $ 9({\bf 22},\1)_{5/3} + 9(\1,{\bf 10})_{-4/3}$ 
\\
& $\frac{1}{3}(1^4,2,0^{11})$ & &$+18(\1,\5)_{-2/3}$ 
\\\hline
3c & $SO(16) \times SU(8) \times U(1)$ & 
$({\bf 16},\8)_1+(\1,{\bf 28})_2+2(\1,\1)_0$
& $9(\1,{\bf 28})_{-2/3}+18(\1,\1)_{8/3}$
\\
& $\frac{1}{3}(1^8,0^8)$ & & 
\\\hline
3d & $SO(10) \times SU(11) \times U(1)$ & 
$({\bf 10},{\bf 11})_1+(\1,{\bf 55})_2+ 2(\1,\1)_0$ &
$9(\1,{\bf 11})_{-8/3}+9(\ov{\bf 16},\1)_{-11/6}$
\\
& $\frac{1}{3}(1^{10},2,0^5)$ & & 
\\\hline
3e & $SU(14) \times SU(2)^2 \times U(1)$ & $({\bf 14},\2,\2)_1+({\bf
  91},\1,\1)_2+2(\1)_0$ 
& $9(\1)_{14/3} + 9({\bf 14},\2,\1)_{-4/3}$
\\
& $\frac{1}{3}(1^{14},0^2)$ & &  $+18(\1,\1,\2)_{-7/3}$ 
\\\hline
\end{tabular}
\end{center}
\caption{Perturbative $SO(32)$ heterotic orbifold spectra on
$T^4/\Z_N$ for $N=2,3$.}
\label{Tab:SO32_Pert_Orbifold_Spectra}
\end{table}

\subsubsection{$T^4/\Z_2$ and $T^4/\Z_3$ models}
Modular invariance forces the number of possible $T^4/\Z_2$ models
to three, with labels 2a,b,c. Similarly the number of $T^4/\Z_3$
models is only five, with labels \mbox{3a - e}. Gauge groups and untwisted
matter can be computed from the shift vectors as discussed above. 
The total spectrum, as already shown in~\cite{Aldazabal:1997wi} except
for the $U(1)$  charges,\footnote{Model 3b including $U(1)$ charges
has been listed before in~\cite{Erler:1993zy}. Our charge
normalization differs and is chosen such that in the untwisted sector
the same charges as for smooth $K3$ embeddings with  $U(4)$ gauge
group occur, i.e. charge 1 for the fundamental and 2 for the
antisymmetric representation of $SU(5)$.} is summarized in
table~\ref{Tab:SO32_Pert_Orbifold_Spectra}.

\begin{table}[t]
\renewcommand{\arraystretch}{1.2}
\begin{center}
\begin{tabular}{|c||c|c|c|c|}
\hline
\hline
$p=n_1+4 n_2$& $p=2$ & $p=10$& $p=18$ & $p=26$\\
\hline\hline
$({\bf 1},{\bf 1},{\bf 2^{n_2-1}_+})_{\bf\frac{n_1}{4}}$ 
& 24 & 12 & 8 & 4 \\
\hline
$({\bf 1},{\bf \ov n_1},{\bf 2^{n_2-1}_-})_{\bf\frac{n_1-4}{4}}$&
12&8&4&0\\\hline
$({\bf 2n_0},{\bf 1},{\bf 2^{n_2-1}_-})_{\bf\frac{n_1}{4}}$&
8&4&0&0\\\hline
$\left({\bf 1},{\bf \frac{\ov n_1(\ov n_1-1)}{2}},
{\bf 2^{n_2-1}_+}\right)_{\bf\frac{n_1-8}{4}}$&
8&4&0&0\\\hline
$({\bf 2n_0},{\bf \ov n_1},{\bf 2^{n_2-1}_+})_{\bf\frac{n_1-4}{4}}$&
4&0&0&0\\\hline
\end{tabular}
\end{center}
\caption{In this table we summarize the $g$-twisted matter content with
{\it vectorial weight} in $SO(32)$ heterotic $T^4/\Z_4$ models with shift
vector ${\bf V}_{\bf a}=\left(0^{n_0=16-n_1-n_2},1^{n_1},2^{n_2}\right)/4$.
We show the multiplicity with which each allowed representation
of the gauge group enters in a model, the latter being
specified by its shift vector ${\bf V}_{\bf a}$ via $p=n_1+4 n_2$.
}
\label{Tab:SO(32)_Z4a}
\end{table}

\subsubsection{$T^4/\Z_4$ models}
\paragraph{Classification of the models:\\}

Two shift vectors are equivalent, i.e. produce the same models, if
their difference can be written in terms of the weight vectors of the
adjoint or spinorial representation of $SO(32)$, up to irrelevant
sign flips. 
This implies that a complete classification is given assuming the 
following ansatz for the shift vectors
\bea
&&
{\bf V_a}=\frac{1}{4}\left(0^{n_0=16-n_1-n_2},1^{n_1},2^{n_{2}}\right),
\\
&&
{\bf V_b}=\frac{1}{4}\left(0^{n_0=16-n_1},1^{n_1-1},3\right).
\eea
Notice that, for $n_2=0$ and fixed value of $n_1$, ${\bf V_a}$ and
${\bf V_b}$ are equivalent for what concerns the gauge symmetry breaking
and the untwisted matter content. Nevertheless, they produce models with
different twisted spectrum \cite{Nilles:2006np}.
The modular invariance condition can be written as
$n_1+4 n_2=2 \; {\rm mod} \; 8$ for both ${\bf V_a}$ and ${\bf V_b}$,
in the latter case with $n_2=0$.
 
For ${\bf V_a}$ the following two options are allowed:
\bea
n_1=2+8p_1, && n_2=2p_2;\\
n_1=6+8p_3, && n_2=2p_4+1.
\eea
The requirement $n_1+n_2\le 16$ constrains the possible values of
$p_i$.
We have (inequivalent) solutions with $n_1=2$, $n_2=0,2,4,6$, and we
label the corresponding models by 4a,b,c,d; $n_1=10$, $n_2=0,2$, with
labels 4e,f; $n_1=6$, $n_2=1,3,5$, with labels 4g,h,i; and
$n_1=14$, $n_2=0$, with label 4j.

For ${\bf V_b}$ we have two models with $n_1=2$ or $n_1=10$, 
4a' and 4e'. These models have the same gauge group
as 4a and 4e, respectively, but have different twisted matter.

For the spinorial shifts, a similar approach can be followed with
shift vector
\bea
&&{\bf V}_{\bf S}=\frac{1}{8}\left(1^{n_1},3^{n_3=16-n_{1}}\right).
\eea
In this case a single parameter is present, and the modular invariance
condition can be rephrased as $n_1=1+4p$. Thus, there are four models
with labels 4k,l,m,n.

\paragraph{Gauge groups and untwisted matter:\\}
Gauge groups and untwisted matter can be deduced from the general
formulae given above.
In the vectorial shift case the gauge group is
$SO(2n_0)\times SU(n_1)\times SO(2n_2)\times U(1)$, with untwisted
matter
\bea
({\bf 2n_0},{\bf n_1},{\bf 1})_1\oplus ({\bf 1},{\bf n_1},{\bf 2n_2})_1
\oplus
({\bf 1},{\bf 1},{\bf 1})_0.
\eea
In the spinorial shift case the gauge group is
$SU(n_1)\times SU(n_3)\times U(1)^2$, with untwisted matter given by 
\bea
\left(\frac{{\bf n_1 (n_1-1)}}{\bf 2},{\bf 1}\right)_{2,0}
\oplus
\left({\bf 1},\frac{{\bf n_2 (n_2-1)}}{\bf 2}\right)_{0,2}
\oplus
({\bf n_1},{\bf \ov n_2})_{1,-1}
\oplus
({\bf 1},{\bf 1})_{0,0}.
\eea

\begin{table}[t]
\renewcommand{\arraystretch}{1.2}
\begin{center}
\begin{tabular}{|c||c|c|c|c|}
\hline
\hline
$q=n_1+4 n_0$& $q=2$ & $q=10$& $q=18$ & $q=26$\\
\hline\hline
$({\bf 2^{n_0-1}_{(-1)^{n_2}}},{\bf 1},{\bf 1})_{\bf-\frac{n_1}{4}}$ &
24 & 12 & 8 & 4 \\
\hline
$({\bf 2^{n_0-1}_{-(-1)^{n_2}}},{\bf n_1},{\bf 1})_{\bf-\frac{n_1-4}{4}}$&
12&8&4&0\\\hline
$({\bf 2^{n_0-1}_{-(-1)^{n_2}}},{\bf 1},{\bf 2n_2})_{\bf-\frac{n_1}{4}}$&
8&4&0&0\\\hline
$\left({\bf 2^{n_0-1}_{(-1)^{n_2}}},{\bf \frac{n_1(n_1-1)}{2}},
{\bf 1}\right)_{\bf-\frac{n_1-8}{4}}$&
8&4&0&0\\\hline
$({\bf 2^{n_0-1}_{(-1)^{n_2}}},{\bf n_1},{\bf 2n_2})_{\bf-\frac{n_1-4}{4}}$&
4&0&0&0\\\hline
\end{tabular}
\end{center}
\caption{In this table we summarize the $g$-twisted matter content
with {\it spinorial weight} vector in $SO(32)$ heterotic $T^4/\Z_4$ models
with shift vectors 
${\bf V}_{\bf a}$.
}
\label{Tab:SO(32)_Z4b}
\end{table}

\begin{table}[t]
\renewcommand{\arraystretch}{1.8}
\begin{center}
\begin{tabular}{|c|c|}
\hline\hline
U &
  $\left({\bf \frac{n_1(n_1-1)}{2}},\1\right)_{2,0}$\hspace{-.4cm}
+ $\left(\1,{\bf\frac{n_3(n_3-1)}{2}}\right)_{0,2}$\hspace{-.4cm}
+ $({\bf n_1},{\bf \ov n_3})_{1,-1}$ + 2 $(\1,\1)_{0,0}$
\\\hline\hline
T &
\begin{tabular}{c|c|c|c|c}
$n_3=$ 
& 3  & 7  & 11 & 15  \\ \hline\hline
$(\1,\1)_{\frac{n_1}{8},\frac{3 n_3}{8}}$
& 12 & 8  & 4  & 0   \\ \hline
$\left(\1,{\bf \frac{\ov n_3 (\ov n_3-1)}{2}}\right)
_{\frac{n_1}{8},\frac{3 n_3-16}{8}}$
& 8  & 4  & 0  & 0   \\ \hline
$({\bf \ov n_1},{\bf\ov n_3})_{\frac{n_1-8}{8},\frac{3 n_3-8}{8}}$
& 4  & 0  & 0  & 0   \\ \hline
$(\1,{\bf \ov n_3})_{-\frac{3n_1}{8},-\frac{n_3+8}{8}}$
& 0  & 0  & 0  & 4   \\ \hline
\mbox{}\hspace{.5cm}
$\left({\bf \frac{n_1(n_1-1)(n_1-2)}{6}},\1\right)
_{-\frac{3n_1-24}{8},-\frac{n_3}{8}}$\hspace{.5cm}
& 0  & 0  & 4  & 8   \\ \hline
$(\1,{\bf n_3})_{-\frac{3n_1}{8},-\frac{n_3-8}{8}}$
& 0  & 0  & 4  & 8   \\ \hline
$({\bf n_1},\1)_{-\frac{3n_1-8}{8},-\frac{n_3}{8}}$
& 0  & 4  & 8  & 12
\end{tabular}
\\\hline\hline
$\rm T^2$ &
6  $({\bf \ov n_1},\1)_{\frac{n_1-4}{4},-\frac{n_3}{4}}$ +
10 $(\1,{\bf n_3})_{\frac{n_1}{4},-\frac{n_3-4}{4}}$ 
\\\hline\hline
\end{tabular}
\end{center}
\caption{Matter content of $SO(32)$ heterotic models on $T^4/\Z_4$
with spinorial shift vector
${\bf V}_{\bf S}=\left(1^{n_1=16-n_3},3^{n_3}\right)/8$.
The $U$ entry summarizes the untwisted spectrum.
The $T$ table summarizes the $g$-twisted spectrum giving
the multiplicity with which each allowed representation
of the gauge group enters in a model, the latter being
specified by its shift vector ${\bf V}_{\bf S}$ via $n_3$.
The $T^2$ line summarizes the $g^2$-twisted spectrum.}
\label{Tab:SO(32)_Z4spi}
\end{table}

\begin{table}[t]
\renewcommand{\arraystretch}{1.2}
\begin{center}
\begin{tabular}{|c||c|c|c|c|}
\hline
\hline
$n_1$& $n_1=2$ & $n_1=6$& $n_1=10$ & $n_1=14$\\
\hline\hline
$({\bf 1},{\bf 2^{n_1-1}_{(-1)^{n_2}}})$
& 4 & 1 & 0 & 0 \\
\hline
$({\bf 2n_0+2n_2},{\bf 2^{n_1-1}_{-(-1)^{n_2}}})$
& 1 & 0 & 0 &0\\\hline
$({\bf 2^{n_0+n_2-1}_{-(-1)^{n_2}}},{\bf 2n_1})$
& 0 & 0 & 0 & 1\\\hline
$({\bf 2^{n_0+n_2-1}_{(-1)^{n_2}}},{\bf 1})$
& 0 & 0 & 1 & 4 
\\\hline
\end{tabular}
\end{center}
\caption{The multiplicity of a given representation of the gauge
group in a model. The gauge group is specified by $n_1$ for the orbifold $T^4/\Z_2$, 
which is relevant as $g^2$ twisted sector in the $T^4/\Z_{4}$ orbifold
(more in general, as $g^N$ twisted sector in any $T^4/\Z_{2N}$ orbifold).}
\label{Tab:SO(32)_Z4double}
\end{table}

\begin{table}[h!]\footnotesize
\renewcommand{\arraystretch}{1.2}
\begin{center}
\begin{tabular}{|c c c|c|}
\hline\hline
$n_1=2$ & & reduced states & $\Z_4$ phase\\\hline
$({\bf 1},{\bf 2_+})$ 
& $\rightarrow$ & 
$({\bf 1},{\bf 1},{\bf 1})_{\bf 1}$ &
$(-1)^{n_2}\times e^{\pi i/2}$\\
&&$({\bf 1},{\bf 1},{\bf 1})_{\bf -1}$ &
$(-1)^{n_2}\times e^{-\pi i/2}$
\\\hline
$({\bf 2n_0+2n_2},{\bf 2_-})$ 
& $\rightarrow$ & 
$({\bf 2n_0},{\bf 2},{\bf 1})_{\bf 0}$ &
$(-1)^{n_2}$\\
&&$({\bf 1},{\bf 2},{\bf 2n_2})_{\bf 0}$ &
$-(-1)^{n_2}$\\\hline\hline
$n_1=6$ & &  reduced states & $\Z_4$ phase \\\hline
$({\bf 1},{\bf 2^5_-})$ 
& $\rightarrow$ & 
$({\bf 1},{\bf \ov 6},{\bf 1})_{\bf 2}$ &
$-(-1)^\frac{n_2+1}{2}$\\
&&$({\bf 1},{\bf 20},{\bf 1})_{\bf 0}$ &
$(-1)^\frac{n_2+1}{2}$\\
&&$({\bf 1},{\bf 6},{\bf 1})_{\bf -2}$ &
$-(-1)^\frac{n_2+1}{2}$
\\\hline\hline
$n_1=10$ & & reduced states  & $\Z_4$ phase \\\hline
$({\bf 2^5_+},{\bf 1})$ 
& $\rightarrow$ & 
$({\bf 2^{n_0-1}_+},{\bf 1},{\bf 2^{n_2-1}_+})_{\bf 0}$ &
$-1$\\
&&
$({\bf 2^{n_0-1}_-},{\bf 1},{\bf 2^{n_2-1}_-})_{\bf 0}$ &
$+1$
\\\hline\hline
$n_1=14$ ($n_2=1$)& & reduced states & $\Z_4$ phase\\\hline
$({\bf 2_+},{\bf 28})$ 
& $\rightarrow$ & 
$({\bf 1},{\bf 14},{\bf 1})_{\bf \frac{1}{2},1,\frac{1}{2}}$ &
$-1$\\
&&
$({\bf 1},{\bf  \ov 14},{\bf 1})_{\bf -\frac{1}{2},-1,-\frac{1}{2}}$ &
$-1$\\
&&
$({\bf 1},{\bf \ov 14},{\bf 1})_{\bf \frac{1}{2},-1,\frac{1}{2}}$ &
$+1$\\
&&
$({\bf 1},{\bf 14},{\bf 1})_{\bf -\frac{1}{2},1,-\frac{1}{2}}$ &
$+1$\\\hline
$({\bf 2_-},{\bf 1})$ 
& $\rightarrow$ & 
$({\bf 1},{\bf 1},{\bf 1})_{\bf \frac{1}{2},0,-\frac{1}{2}}$ &
$e^{-\pi i/2}$\\
&&
$({\bf 1},{\bf  1},{\bf 1})_{\bf -\frac{1}{2},0,\frac{1}{2}}$ &
$e^{\pi i/2}$
\\\hline\hline
\end{tabular}
\end{center}
\caption{The $\Z_4$ reduction of the $T^4/\Z_2$ twisted states.
Each state should be multiplied by the multiplicity given in the
table~5,
remembering that a multiplicity $1$ brings no extra
bosonic phases, while a multiplicity $4$ must be split into a $2$
with $e^{\pi i/2}$ phase and a $2$ with $e^{-\pi i/2}$ phase.
The states with global phase +1 then receive a 10/2 multiplicity
from the degeneracy of the fixed points, those with phase -1 receive
instead a 6/2 multiplicity.}
\label{Tab:SO(32)_Z4red}
\end{table}

\paragraph{$g$-Twisted matter:\\}
The $g$-twisted matter can be computed by re-quantizing the closed
heterotic string with twisted boundary conditions.
This boils down to the identification of the vacua of the sector
(i.e. in which representation of the gauge group and of the internal
holonomy group they are) and to the classification of the excitations
that lift them to the zero-mass level (indeed, all the tachyonic vacua
are projected out of the spectrum).

There are two possible vacua, one with vectorial weight vector,
the other with spinorial weight vector. Given a model with shift
vector\footnote{We do not
discuss here the two models with vector of the form ${\bf V_b}$,
their massless spectra are directly given in
table~\ref{Tab:SO(32)_Z4allinall}. The models with spinorial shift
vector ${\bf V}_{\bf S}$ are discussed later.}
\bea
{\bf V_a}=\frac{1}{4}\left(0^{n_0=16-n_1-n_2},1^{n_1},2^{n_2}\right)
\eea
and unbroken gauge group
\bea
SO(2n_0)\times U(n_1)\times SO(2 n_2),
\eea
the two vacua are in the 
$({\bf 1},{\bf 1},{\bf 2^{n_2-1}_+})_{\bf\frac{n_1}{4}}$
and in the
$({\bf 2^{n_0-1}_{(-1)^{n_2}}},{\bf 1},{\bf 1})_{\bf-\frac{n_1}{4}}$ 
representations, respectively.

For each of the two vacua,
the excitations can be switched on among two classes: On the one hand
they can change the gauge representation of the vacuum, and then we
call them ``gauge'' excitations. On the other hand they can change
the holonomy representation of the vacuum, if we consider excitations
due to the four $X^m$ target space bosons along compact directions, 
and then we call them 
``bosonic'' excitations. The excitations of the second class are 
``visible'' from the six dimensional perspective since they change
the degeneration of the corresponding twisted state.
Both kinds of excitations produce a mass lift, and in a model each
gauge excitation is matched with a bosonic excitation to produce a
massless state. Since no negative lifting is present, only a few
excitations that do not overclose the zero mass condition are allowed.
All others produce massive string modes.

Thus, for each vacuum we can just list the ``allowed'' excited
gauge representations, and the allowed multiplicities coming from the
bosonic excitations. Then, each representation is present in a model with
the multiplicity such that the state is massless. In case the state
is always massive, we can just think that the multiplicity is zero.
The matching depends on the energy of the vacuum, given by the
shift vector ${\bf V}_{\bf a}$ via the numbers $n_0$, $n_1$ and $n_2$.

In our specific case, the possible representations due to the vacuum with
vectorial weight are listed in table \ref{Tab:SO(32)_Z4a}, those
due to the vacuum with spinorial weight are listed in table 
\ref{Tab:SO(32)_Z4b}. The possible degeneracies due to oscillators are $1,2,3,6$,
that must be multiplied by $4$, the number of fixed points.
In  tables \ref{Tab:SO(32)_Z4a} and  \ref{Tab:SO(32)_Z4b}
we also resume the matching between degeneracies and representations in the
models: the degeneracy with which a representation enters in a model
is given as a function of the characteristic numbers 
$n_0$, $n_1$ and $n_2$ of the model.

As last remarks, we mention some caveats that should be taken
into account when reading the tables.
In case $p=n_1+4n_2>14$, the related column is not listed in
table~\ref{Tab:SO(32)_Z4a}, since all the entries are 0. Similarly
for $q=n_1+4 n_0>14$ in table~\ref{Tab:SO(32)_Z4b}.

In case $n_0=1$ and/or $n_2=1$ the gauge group $SO(2n_0)$ and/or
$SO(2n_2)$ reduces to two/one extra $U(1)$ factor(s), and extra $U(1)$ charges
are present. The latter are as follows. The
${\bf 2^{n_i-1}_\pm}$ representation,  a spinorial representation of
$SO(2n_i)$ with $\pm$ chirality,  shrinks to a singlet of  $U(1)$ with 
charge $\pm 1/2$; the ${\bf 2n_i}$ representation instead shrinks to
two singlets with charges $\pm 1$.
In case $n_i=0$ ($i=0,2$), the representations including ${\bf 2n_i}$
or ${\bf 1_i}$ or ${\bf 2^{n_i-1}_+}$ are still present, but the
${\bf 2n_i}$ or ${\bf 1_i}$ or ${\bf 2^{n_i-1}_+}$ entry, related to
an inexistent $SO(0)$ group, drops. 
The representations including  ${\bf 2^{n_i-1}_-}$ are instead removed
from the spectrum by the GSO projection.
As an example of the latter caveat, consider the case $n_1=2$, $n_2=0$
(``standard embedding'').
From table~\ref{Tab:SO(32)_Z4a} we read off the matter content with
vectorial weight from the column with $p=n_1+4 n_2=2$.
Since $n_2=0$ we have to keep all the states in spinorial
representations with $+$  chirality and remove those with $-$
chirality.
Thus we have the following matter content: (i) 24 multiplets in the
$({\bf 1},{\bf 1})_{\bf \frac{1}{2}}$, (ii) eight multiplets in the
$({\bf 1},{\bf 1})_{\bf -\frac{3}{2}}$ and (iii) four in the
$({\bf 28},{\bf 2})_{\bf -\frac{1}{2}}$ representations of the gauge
group $SO(28) \times SU(2) \times U(1)$. 
From table~\ref{Tab:SO(32)_Z4b}, instead, we get no extra state,
since $q=n_1+4 n_0>14$.

In the case of the spinorial shifts we use the same approach as explained
above. The results are listed in table~\ref{Tab:SO(32)_Z4spi}.

\paragraph{$g^2$-twisted matter:\\}
The second twisted sector is built starting from $2{\bf V_a}=\frac{1}{4}
(0^{n_0},2^{n_1},4^{n_2})\sim\frac{1}{2}(0^{n_0+n_2},1^{n_1})$,
i.e. from the twisted sector of the $T^4/\Z_2$ orbifold. Given this,
we have to orbifold such a sector by projecting out the states that
are not invariant under the $\Z_4$ operator.
We can  split the computation into two steps:
In the first we compute the twisted sector of $T^4/\Z_2$,
summarized in table~\ref{Tab:SO(32)_Z4double}, then we project it.
The $SO(32)$ gauge symmetry is broken by $\Z_2$ to $SO(2n_0+2n_2)
\times SO(2n_1)$, and thus, before the $\Z_4$ projection, the twisted
states are organized in multiplets of such a symmetry. We
have a multiplicity of one for states without bosonic excitations,
and of four for those with an excitation. On top of this, the
multiplicity from the fixed point degeneracy, $16$, should be added.
We do not include the latter in table~\ref{Tab:SO(32)_Z4double}.
The table is built as previously, the characteristic number is now
$n_1$. We resume in it all  states.
The space-time multiplicity is, as in the $g^1$-twisted sector,
just a factor of two, but in this case we do not have a $g^3$-twisted
sector rising such a multiplicity to four, i.e. the multiplicity of
a full hyper multiplet. Then, when reading table~\ref{Tab:SO(32)_Z4double},
we have to remember the factor 16 from the fixed points, but also
a factor 1/2 necessary in order to correctly count the number of
hyper multiplets.

The second step is the orbifold projection including the $\Z_4$ phases
of the obtained states. We summarize the reduction of the states in 
table~\ref{Tab:SO(32)_Z4red}. Notice that the phase must be combined
with the one coming from the bosonic excitations; it is $+1$ in
case of no excitation, and it is $e^{\pi i/2}$ for a doublet of the 
four excited states and $e^{-\pi i/2}$ for the other doublet.
The total phase is always either $+1$ or $-1$.
In case the phase is $+1$ the states receive an extra 10/2=5 multiplicity
from the fixed point degeneracy, if instead the phase is $-1$ the
multiplicity is 6/2=3.

In the case of the spinorial shifts we use the same approach as explained
above, the results are listed in table~\ref{Tab:SO(32)_Z4spi}.

The complete massless spectra for heterotic $SO(32)$ orbifolds
on $T^4/\Z_4$ are given in table~\ref{Tab:SO(32)_Z4allinall} for the
vectorial and table~\ref{Tab:SO(32)_Z4spiallinall} for the spinorial
shifts.

\subsection{$E_8 \times E_8$ models}
\subsubsection{$T^4/\Z_2$ and $T^4/\Z_3$  models}
These models have been classified, except for the $U(1)$ charge
assignments, in~\cite{Aldazabal:1997wi}. We summarize them in
table~\ref{Tab:E82_Pert_Orbifold_Spectra}.
The standard embeddings IIa, IIIa (with different $U(1)$ charge
normalization)  as well as the cases IIIc, IIIe were already presented
in~\cite{Erler:1993zy}. Model IIIe is also discussed
in some detail in~\cite{Kaplunovsky:1999ia}, IIId has been computed
before in~\cite{Aldazabal:1995yw}.

\begin{table}[p]
\renewcommand{\arraystretch}{1.2}
\begin{center}\footnotesize
\begin{tabular}{|c||l||c|}
\hline\hline
\# \& Gauge group & & Matter
\\\hline\hline
4a  & $U$ & 
   $({\bf 28},{\bf 2})_{\bf 1}$ +
2  $(\1,\1)_{\bf 0}$
\\
$SO(28)\times SU(2) \times U(1)$ &$T$& 
24 $(\1,\1)_{\bf \frac{1}{2}}$ +
8  $(\1,\1)_{\bf -\frac{3}{2}}$ +
4  $({\bf 28},{\bf 2})_{\bf -\frac{1}{2}}$
\\
(s. e.) $\frac{1}{4}\left(0^{14},1^{2}\right)$ & $T^2$ &
32  $(\1,\1)_{\bf  1}$ +
5   $({\bf 28},{\bf 2})_{\bf 0}$
\\\hline\hline
4a'  & $U$ & 
   $({\bf 28},{\bf 2})_{\bf 1}$ +
2  $(\1,\1)_{\bf 0}$
\\
$SO(28)\times SU(2) \times U(1)$ &$T$& 
12 $(\1,{\bf 2})_{\bf -\frac{1}{2}}$ +
4  $(\1,{\bf 2})_{\bf \frac{3}{2}}$ +
8  $({\bf 28},{\bf 1})_{\bf \frac{1}{2}}$
\\
 $\frac{1}{4}\left(0^{14},1,3\right)$ & $T^2$ &
32  $(\1,\1)_{\bf 1}$ +
5   $({\bf 28},{\bf 2})_{\bf 0}$
\\\hline\hline
4b  & $U$ & 
   $({\bf 1},{\bf 2},{\bf 4})_{\bf 1}$ + 
   $({\bf 24},{\bf 2},\1)_{\bf 1}$ +
2  $(\1,\1,\1)_{\bf 0}$
\\
$SO(24)\times SU(2)\times SO(4) \times U(1)$ &$T$&
$\begin{array}{c} 
12  (\1,\1,{\bf 2_+})_{\bf \frac{1}{2}} +
8  (\1, {\bf 2}, {\bf 2_-})_{\bf -\frac{1}{2}} +\\
4  ({\bf 24},\1,{\bf 2_-})_{\bf \frac{1}{2}} +
4  (\1, {\bf 1}, {\bf 2_+})_{\bf -\frac{3}{2}}
\end{array}$
\\
$\frac{1}{4}\left(0^{12},1^{2},2^{2}\right)$
& $T^2$ &
32 $(\1,\1,\1)_{\bf  1}$ +
5  $(\1,{\bf 2},{\bf 4})_{\bf 0}$ +
3  $({\bf 24},{\bf 2},\1)_{\bf 0}$
\\\hline\hline
4c  & $U$ & 
   $({\bf 1},{\bf 2},{\bf 8})_{\bf 1}$ + 
   $({\bf 20},{\bf 2},\1)_{\bf 1}$ +
2  $(\1,\1,\1)_{\bf 0}$
\\
$SO(20)\times SU(2)\times SO(8) \times U(1)$ &$T$&
8  $(\1,\1,{\bf 8_+})_{\bf \frac{1}{2}}$ +
4  $(\1, {\bf 2}, {\bf 8_-})_{\bf -\frac{1}{2}}$
\\
$\frac{1}{4}\left(0^{10},1^{2},2^{4}\right)$
& $T^2$ &
32 $(\1,\1,\1)_{\bf  1}$ +
3  $(\1,{\bf 2},{\bf 8})_{\bf 0}$ +
5  $({\bf 20},{\bf 2},\1)_{\bf 0}$
\\\hline\hline
4d  & $U$ & 
   $({\bf 1},{\bf 2},{\bf 12})_{\bf 1}$ + 
   $({\bf 16},{\bf 2},\1)_{\bf 1}$ +
2  $(\1,\1,\1)_{\bf 0}$
\\
$SO(16)\times SU(2)\times SO(12) \times U(1)$ &$T$&
4  $(\1, \1, {\bf 32_+})_{\bf \frac{1}{2}}$
\\
$\frac{1}{4}\left(0^{8},1^{2},2^{6}\right)$
& $T^2$ &
32 $(\1,\1,\1)_{\bf  1}$ +
5  $(\1,{\bf 2},{\bf 12})_{\bf 0}$ +
3  $({\bf 16},{\bf 2},\1)_{\bf 0}$
\\\hline\hline
4e  & $U$ & 
   $({\bf 12},{\bf 10})_{\bf 1}$ +
2  $(\1,\1)_{\bf 0}$
\\
$SO(12)\times SU(10) \times U(1)$ &$T$&
12  $(\1, \1)_{\bf \frac{5}{2}}$ +
4   $(\1,{\bf \ov{45}})_{\bf\frac{1}{2}}$
\\
$\frac{1}{4}\left(0^{6},1^{10}\right)$
& $T^2$ &
3  $({\bf 32_+},\1)_{\bf 0}$
\\\hline\hline
4e'  & $U$ & 
   $({\bf 12},{\bf 10})_{\bf 1}$ +
2  $(\1,\1)_{\bf 0}$
\\
$SO(12)\times SU(10) \times U(1)$ &$T$&
8   $(\1, {\bf \ov{10}})_{\bf \frac{3}{2}}$ +
4   $({\bf 12},\1)_{\bf\frac{5}{2}}$
\\
$\frac{1}{4}\left(0^{6},1^{9},3\right)$
& $T^2$ &
5  $({\bf 32_+},\1)_{\bf 0}$
\\\hline\hline
4f  & $U$ & 
   $({\bf 1},{\bf 10},{\bf 4})_{\bf 1}$ + 
   $({\bf 8},{\bf 10},\1)_{\bf 1}$ +
2  $(\1,\1,\1)_{\bf 0}$
\\
$SO(8)\times SU(10)\times SO(4) \times U(1)$ &$T$&
8  $(\1, \1, {\bf 2_+})_{\bf \frac{5}{2}}$ +
4  $(\1, {\bf \ov{10}}, {\bf 2_-})_{\bf \frac{3}{2}}$ +
4  $({\bf 8_+}, \1, \1)_{\bf -\frac{5}{2}}$
\\
$\frac{1}{4}\left(0^{4},1^{10},2^{2}\right)$
& $T^2$ &
3  $({\bf 8_+},\1,{\bf 2_+})_{\bf 0}$ +
5  $({\bf 8_-},\1,{\bf 2_-})_{\bf 0}$
\\\hline\hline
4g  & $U$ & 
   $({\bf 1},{\bf 6})_{\bf 1, 1}$ + 
   $({\bf 1},{\bf 6})_{\bf 1,- 1}$ + 
   $({\bf 18},{\bf 6})_{\bf 1,0}$ +
2  $(\1,\1)_{\bf 0 ,0}$
\\
$SO(18)\times SU(6)\times U(1)^2$ &$T$& 
12 $(\1,\1)_{\bf \frac{3}{2},\frac{1}{2}}$ + 
8  $(\1, {\bf\ov 6})_{\bf \frac{1}{2},-\frac{1}{2}}$ +
4  $({\bf 18},\1)_{\bf \frac{3}{2},-\frac{1}{2}}$ +
4  $(\1,{\bf \ov{15}})_{\bf -\frac{1}{2},\frac{1}{2}}$ \\
$\frac{1}{4}\left(0^{9},1^{6},2\right)$ & $T^2$ &
3  $(\1,{\bf 20})_{\bf 0,0}$ +
10  $(\1,{\bf 6})_{\bf -2,0}$
\\\hline\hline
4h  & $U$ & 
   $({\bf 1},{\bf 6},{\bf 6})_{\bf 1}$ + 
   $({\bf 14},{\bf 6},\1)_{\bf 1}$ +
2  $(\1,\1,\1)_{\bf 0}$
\\
$SO(14)\times SU(6)\times SO(6) \times U(1)$ &$T$& 
8  $(\1,\1,{\bf 4})_{\bf \frac{3}{2}}$ +
4  $(\1, {\bf\ov 6},{\bf\ov 4})_{\bf \frac{1}{2}}$ 
\\
$\frac{1}{4}\left(0^{7},1^{6},2^{3}\right)$
& $T^2$ &
5  $(\1,{\bf 20},\1)_{\bf 0}$ +
6  $(\1,{\bf 6},\1)_{\bf -2}$
\\\hline\hline
4i  & $U$ & 
   $({\bf 1},{\bf 6},{\bf 10})_{\bf 1}$ + 
   $({\bf 10},{\bf 6},\1)_{\bf 1}$ +
2  $(\1,\1,\1)_{\bf 0}$
\\
$SO(10)\times SU(6)\times SO(10) \times U(1)$ &$T$& 
4  $(\1,\1,{\bf 16})_{\bf \frac{3}{2}}$ +
4  $(\ov{\bf 16},\1, \1)_{\bf -\frac{3}{2}}$ 
\\
$\frac{1}{4}\left(0^{5},1^{6},2^{5}\right)$
& $T^2$ &
3  $(\1,{\bf 20},\1)_{\bf 0}$ +
10  $(\1,{\bf 6},\1)_{\bf -2}$
\\\hline\hline
4j  & $U$ & 
   $({\bf 14})_{\bf 0, 1, 1}$ +
   $({\bf 14})_{\bf 0, 1,- 1}$ +
   $({\bf 14})_{\bf  1, 1,0}$ +
   $({\bf 14})_{\bf - 1, 1,0}$ +
2  $(\1)_{\bf 0,0,0}$
\\
$SU(14)\times U(1)^3$ &$T$& 
8  $({\bf 1})_{\bf 0, \frac{7}{2}, \frac{1}{2}}$ +
8  $({\bf 1})_{\bf -\frac{1}{2}, -\frac{7}{2},0}$ +   
4  $({\bf\ov{14}})_{\bf 0, \frac{5}{2},-\frac{1}{2}}$ +
4  $({\bf 14})_{\bf \frac{1}{2}, -\frac{5}{2},0}$
\\
$\frac{1}{4}\left(0,1^{14},2\right)$
& $T^2$ &
$\begin{array}{c}
32   ({\bf 1})_{\bf \frac{1}{2}, 0, -\frac{1}{2}} +
10   ({\bf 14})_{\bf -\frac{1}{2},1, -\frac{1}{2}} +
6   ({\bf 14})_{\bf \frac{1}{2},1, \frac{1}{2}}
\end{array}$
\\\hline\hline
\end{tabular}
\end{center}
\caption{Gauge group and massless matter of $SO(32)$ heterotic models
on $T^4/\Z_4$ with vectorial shift vector.}
\label{Tab:SO(32)_Z4allinall}
\end{table}

\begin{table}[t]
\renewcommand{\arraystretch}{1.2}
\begin{center}\footnotesize
\begin{tabular}{|c||l||c|}
\hline\hline
\# \& Gauge group & & Matter\\\hline\hline
4k  & $U$ & 
   $({\bf 105})_{\bf 0,2}$ + 
   $({\bf\ov{15}})_{\bf 1, -1}$ + 
2  $(\1)_{\bf 0 ,0}$
\\
$SU(15) \times U(1)^2$ &$T$& 
12 $(\1)_{\bf \frac{5}{8},-\frac{15}{8}}$ + 
8  $({\bf 15})_{\bf -\frac{3}{8},-\frac{7}{8}}$ +
4  $({\bf \ov{15}})_{\bf -\frac{3}{8},-\frac{23}{8}}$ \\
$\frac{1}{8}\left(1,3^{15}\right)$ & $T^2$ &
6  $(\1)_{\bf -\frac{3}{4},-\frac{15}{4}}$ +
10 $({\bf 15})_{\bf \frac{1}{4},-\frac{11}{4}}$
\\\hline\hline
4l  & $U$ & 
   $({\bf 10},\1)_{\bf 2,0}$ + 
   $(\1,{\bf 55})_{\bf 0,2}$ +
   $({\bf 5},{\bf \ov{11}})_{\bf 1,-1}$ + 
2  $(\1,\1)_{\bf 0,0}$
\\
$SU(5)\times SU(11) \times U(1)^2$ &$T$& 
8  $({\bf 5},\1)_{\bf -\frac{7}{8},-\frac{11}{8}}$ +
4  $(\1,{\bf 11})_{\bf -\frac{15}{8},-\frac{3}{8}}$ +
4  $({\bf\ov{10},\1})_{\frac{9}{8},-\frac{11}{8}}$ +
4  $(\1,\1)_{\frac{5}{8},\frac{33}{8}}$
\\
$\frac{1}{8}\left(1^{5},3^{11}\right)$
& $T^2$ &
6  $({\bf \ov{5}},\1)_{\bf \frac{1}{4},-\frac{11}{4}}$ +
10 $(\1,{\bf 11})_{\bf \frac{5}{4},-\frac{7}{4}}$
\\\hline\hline
4m  & $U$ & 
   $({\bf 36},\1)_{\bf 2,0}$ + 
   $(\1,{\bf 21})_{\bf 0,2}$ +
   $({\bf 9},{\bf \ov{7}})_{\bf 1,-1}$ + 
2  $(\1,\1)_{\bf 0,0}$
\\
$SU(9)\times SU(7) \times U(1)^2$ &$T$& 
4  $({\bf 9},\1)_{\bf -\frac{19}{8},-\frac{7}{8}}$ +
4  $(\1,{\bf \ov{21}})_{\bf \frac{9}{8},\frac{5}{8}}$ +
8  $(\1,\1)_{\frac{9}{8},\frac{21}{8}}$
\\
$\frac{1}{8}\left(1^{9},3^{7}\right)$
& $T^2$ &
6  $({\bf \ov{9}},\1)_{\bf \frac{5}{4},-\frac{7}{4}}$ +
10 $(\1,{\bf 7})_{\bf \frac{9}{4},-\frac{3}{4}}$
\\\hline\hline
4n  & $U$ & 
   $({\bf 78},\1)_{\bf 2,0}$ + 
   $(\1,{\bf \ov 3})_{\bf 0,2}$ +
   $({\bf 13},{\bf \ov{3}})_{\bf 1,-1}$ + 
2  $(\1,\1)_{\bf 0,0}$
\\
$SU(13)\times SU(3) \times U(1)^2$ &$T$& 
4  $({\bf \ov{13}},{\bf \ov 3})_{\bf \frac{5}{8},\frac{1}{8}}$ +
8  $(\1,{\bf 3})_{\bf \frac{13}{8},-\frac{7}{8}}$ +
12  $(\1,\1)_{\bf \frac{13}{8},\frac{9}{8}}$
\\
$\frac{1}{8}\left(1^{13},3^3\right)$
& $T^2$ &
6  $({\bf \ov{13}},\1)_{\bf \frac{9}{4},-\frac{3}{4}}$ +
10 $(\1,{\bf 3})_{\bf \frac{13}{4},\frac{1}{4}}$
\\\hline\hline
\end{tabular}
\end{center}
\caption{$SO(32)$ heterotic models on $T^4/\Z_4$ with
spinorial shift vector.}
\label{Tab:SO(32)_Z4spiallinall}
\end{table}

\begin{table}[h!]
\renewcommand{\arraystretch}{1.2}
\begin{center}\footnotesize
\begin{tabular}{|c||c||c|c|}
\hline
\hline
\# & Group & Untwisted & Twisted \\
& Shift & Matter & Matter
\\\hline\hline
IIa & $E_7 \times SU(2) \times E_8$ & 
$({\bf 56},\2)+4(\1,\1)$ & $8({\bf 56},\1)+32(\1,\2)$
\\
& $\half(1^2,0^6) \times (0^8)$ &&
\\\hline
IIb & $SO(16) \times E_7 \times SU(2)$ & 
$(\1;{\bf 56},\2)+({\bf 128};\1,\1)+4(\1)$ & $8({\bf 16},\1,\2)$
\\
& $\half(1,0^7) \times(1^2,0^6)$ & &
\\\hline\hline
IIIa & $E_7 \times U(1) \times E_8$ & 
$({\bf 56},\1)_1+(\1,\1)_2 +2(\1,\1)_0$
& $9({\bf 56},\1)_{\frac{1}{3}}+45 (\1,\1)_{\frac{2}{3}}$  
\\
& $\frac{1}{3}(1^2,0^6)  \times (0^8)$ & & $+ 18(\1,\1)_{\frac{4}{3}}$
\\\hline
IIIb & $SO(14)^2 \times U(1)^2$ & 
$({\bf 14},\1)_{1,0}+(\1,{\bf 14})_{0,1} + 
({\bf 64},\1)_{\frac{1}{2},0} $
& 
$9({\bf 14},\1)_{-\frac{1}{3},\frac{2}{3}}+ 
9(\1,{\bf 14})_{\frac{2}{3},-\frac{1}{3}} $
\\
&$\frac{1}{3}(2,0^7) \times \frac{1}{3}(2,0^7)$ & 
$+ (\1,{\bf 64})_{0,\frac{1}{2}} + 2(\1,\1)_{0,0}$
&$+ 18(\1,\1)_{\frac{2}{3},\frac{2}{3}}$
\\\hline
IIIc & $SU(9) \times E_8$ & $({\bf 84},\1)+2(\1,\1)$ &
$9({\bf 36},\1)+18({\bf 9},\1)$
\\
& $\frac{1}{3}(1^4,2,0^3) \times (0^8)$ & & 
\\\hline
IIId & $E_6 \times SU(3) \times E_7 \times U(1)$ & $({\bf
  27},\3,\1)_0+(\1,\1,{\bf 56})_1$ 
& $9({\bf 27},\1,\1)_{\frac{2}{3}} + 
9(\1,\3,\1)_{\frac{4}{3}}$ 
\\
&  $\frac{1}{3}(1^2,2,0^5)\times \frac{1}{3}(1^2,0^6)$ 
& $+2(\1)_0+(\1)_2$
& $ +18 (\1,\3,\1)_{-\frac{2}{3}}$
\\\hline
IIIe & $ SU(9) \times E_6 \times SU(3)$ & $({\bf 84},\1,\1) +
(\1,{\bf 27},\3) + 2(\1)$ & $9({\bf 9},\1,\3)$ 
\\
&  $\frac{1}{3}(1^4,2,0^3) \times \frac{1}{3}(1^2,2,0^5)$ & & 
\\\hline
\end{tabular}
\end{center}
\caption{Perturbative $E_8 \times E_8$ heterotic orbifold spectra on
$T^4/\Z_N$ for $N=2,3$.}
\label{Tab:E82_Pert_Orbifold_Spectra}
\end{table}

\begin{table}[ht]
\renewcommand{\arraystretch}{1.2}
\footnotesize\mbox{}\hspace{-1.1cm}
\begin{tabular}{|c||c|c||c|c|}
\hline
\hline
Shift & 
$SO(16)$ breaking & 
Untwisted matter & 
$E_8$ enhancement &
Untwisted matter\\
\hline\hline
$\frac{1}{4} \left(0^8\right)$ & 
$SO(16)$ &  -
& 
$E_8$ & -
\\\hline
$\frac{1}{4} \left(4,0^7\right)$ & 
$SO(16)$ & -   
& 
$SO(16)$ & - 
\\\hline
$\frac{1}{4} \left(2,0^7\right)$ & 
$SO(14)\times U(1)$ &  
$({\bf 64})_{\frac{1}{2}}$& 
$SO(14)\times U(1)$ &  
$({\bf 64})_{\frac{1}{2}}$
\\\hline
$\frac{1}{4} \left(1^2,0^6\right)$ & 
$SO(12)\times SU(2)\times U(1)$ &
$({\bf 12},\2)_1 + ({\bf 32_+},\1)_1$ &
$E_7\times U(1)$ &
$({\bf 56},\1)_1$
\\\hline
$\frac{1}{4} \left(1,3,0^6\right)$ & 
$SO(12)\times SU(2)\times U(1)$ &
$({\bf 12},\2)_1 + ({\bf 32_+},\1)_1$ &
$\!\!\!SO(12)\times SU(2)\times U(1)\!\!\!$ &
$({\bf 12},\2)_1 + ({\bf 32_+},\1)_1$
\\\hline
$\frac{1}{4} \left(2^2,0^6\right)$ & 
$SO(12)\times SO(4)$ & - 
&
$E_7\times SU(2)$ & - 
\\\hline
$\frac{1}{4} \left(1^2,2,0^5\right)$ & 
$SO(10)\times SU(2)\times U(1)^2$ &
$\!\!\!\begin{array}{c}
({\bf 10},\2)_{-2,-1} + ({\bf 16},\2)_{1,-1} \\
+ (\1,\2)_{4,-1} + (\1,\2)_{0,3}
\end{array}\!\!\!$
&
$E_6\times SU(2)\times U(1)$ &
$({\bf 27},\2)_{-1} + (\1,\2)_3$
\\\hline
$\frac{1}{4} \left(2^3,0^5\right)$ & 
$SO(10)\times SO(6)$ &
$({\bf 16},\4)$&
$SO(10)\times SO(6)$ &
$({\bf 16},\4)$
\\\hline
$\frac{1}{4} \left(1^5,3,0^2\right)$ & 
$\!\!\!SU(6)\times SU(2)^2\times U(1)\!\!\!$ &
$\!\!\!\begin{array}{c}
({\bf 6},\2,\2)_{-1} + ({\bf 15},\1,\2)_1 \\
+ (\1,\1,\2)_{-3}
\end{array}\!\!\!$
&
$SU(8)\times SU(2)$ &
$({\bf 28},\2)$
\\\hline
$\frac{1}{4} \left(1^7,-1\right)$ & 
$SU(8)\times U(1)$ &
$({\bf 8})_{-3} + ({\bf 56})_{-1}$
&
$SU(8)\times U(1)$ &
$({\bf 8})_{-3} + ({\bf 56})_{-1}$
\\\hline
\end{tabular}
\caption{Gauge groups and untwisted matter for the ten inequivalent $\Z_4$ shift vectors
of $E_8$.
In the second and third column, the gauge groups and untwisted matter obtained along the
same lines as for the 
$SO(32)$ cases are given, and in the last two columns the enhancements due to the breaking of 
the ${\bf 128}$ gauge boson are listed.
\label{Tab:E8_Z4_shifts}}
\end{table}

\subsubsection{$T^4/\Z_4$ models}
The inequivalent shift vectors for this orbifold background have been
classified  in~\cite{Stieberger:1998yi} as well as the gauge groups,
and the instanton numbers $(k_1,k_2)$ with $k_1+k_2=24$ have been computed for each case.
We briefly review the classification here and give some of the details of the computation
of the massless spectra - which is done here for the first time - for each shift vector. The results are summarized
in table \ref{Tab:E8_Z4_Spectra}.

\paragraph{Gauge symmetry breaking and untwisted matter\\}
We can consider each $E_8$ factor separately. Thus, we split
the shift vector into two subvectors with eight entries each.
As discussed in~\cite{Stieberger:1998yi}, the possible subvectors
are only ten, listed in table \ref{Tab:E8_Z4_shifts}.

The gauge symmetry breaking can then be studied by considering
first the subgroup $SO(16)$ of $E_8$.
The vectors listed in table \ref{Tab:E8_Z4_shifts} are all of the form
\bea
V_{E_8}=\frac{1}{4} \left(0^{n_0},1^{n_1},-1^{n_{-1}},2^{n_2},3^{n_3},4^{n_4}\right),
\eea
and the breaking of $SO(16)$ is
$SO(16)\rightarrow SO(2n_0+2n_4)\times U(n_1+n_{-1}+n_3)\times SO(2n_2)$.

Given the breaking of the $SO(16)$ subgroup, we pass to its enhancement to $E_8$,
due to the presence of gauge bosons in the ${\bf 128}$ representation of $SO(16)$.
In case some of the states in the ${\bf 128}$ representation are left invariant
by the orbifold action, the gauge group is enhanced, as
shown in table \ref{Tab:E8_Z4_shifts}.

The untwisted matter can then be obtained as in the $SO(32)$ case,
with the caveat that now also the ${\bf 128}$ representation is present,
and that, in presence of a gauge enhancement, the untwisted states are
rearranged in multiplets of the enhanced gauge group, as shown in
table \ref{Tab:E8_Z4_shifts}.

The actual $E_8\times E_8$ shift vectors come from a combination
of two $E_8$ shift subvectors.
The possible combinations, allowed by modular invariance, 
have been classified in~\cite{Stieberger:1998yi} and 
are listed in table \ref{Tab:E8_Z4_Spectra}, as well as the
gauge symmetry breaking and matter content which is computed here for the first time.

\begin{table}[p]
\renewcommand{\arraystretch}{1.2}
\footnotesize
\mbox{}\vspace{-.2cm}
\hspace{-.8cm}
\begin{tabular}{|c||c||c|c|}
\hline
\hline
 Gauge group \& Shift & & Matter
\\\hline\hline 
IVa
& $U$ & 
   $({\bf 56};\1)_1$ + 
2  $(\1;\1)_0$
\\
 $E_7 \times U(1) \times E_8$ 
& $T$ & 
4  $({\bf 56};\1)_{-\frac{1}{2}}$ + 
8  $(\1;\1)_{\frac{3}{2}}$ + 
24 $(\1;\1)_{\frac{1}{2}}$
\\
 $\frac{1}{4}(1^2,0^6;0^8)$
& $\!\!T^2\!\!$ & 
   $5({\bf 56};\1)_0$ + 
32 $(\1;\1)_1$ 
\\\hline\hline 
IVb &  $U$ & 
   $({\bf 56};\1,\1)_1$ + 
2  $(\1;\1,\1)_0$ 
\\
$E_7 \times U(1) \times E_7 \times SU(2)$ 
& $T$ & 
12 $(\1;\1,\2)_{\frac{1}{2}}$ +
4  $(\1;\1,\2)_{-\frac{3}{2}}$ +
4  $(\1;{\bf 56},\1)_{\frac{1}{2}}$
\\
 $\frac{1}{4}(1^2,0^6;2^2,0^6)$
& $\!\!T^2\!\!$ & 
32 $(\1;\1,\1)_{1}$ +
3  $({\bf 56};\1,\1)_{0}$
\\\hline\hline 
IVc & U &  
   $({\bf 56};\1)_1$ + 
2  $(\1;\1)_0$
\\
$E_7 \times U(1) \times SO(16)$ 
& $T$ & 
8  $(\1;{\bf 16})_{\frac{1}{2}}$
\\
$\frac{1}{4}(1^2,0^6;4,0^7)$
& $\!\!T^2\!\!$ &
32 $(\1;\1)_{1}$ +
5  $({\bf 56};\1)_0$
\\\hline\hline 
IVd & $U$ &  
   $({\bf 56};\1)_{1,0}$ + 
   $(\1,\8)_{0,-3}$ +
   $(\1,{\bf 56})_{0,-1}$ + 
2  $(\1,\1)_{0,0}$
\\
 $E_7 \times U(1) \times SU(8) \times U(1)$ 
& $T$ &
12 $(\1;\1)_{\frac{1}{2},2}$ +
4  $(\1;\1)_{-\frac{3}{2},2}$ +
4  $(\1; {\bf \ov{28}})_{\frac{1}{2},0}$ +
8  $(\1;{\bf 8})_{\frac{1}{2},-1}$
\\
 $\frac{1}{4}(1^2,0^6;1^7,-1)$
& $\!\!T^2\!\!$ & 
10 $(\1;{\bf 8})_{1,1}$ +
6  $(\1;{\bf 8})_{-1,1}$
\\\hline\hline 
IVe 
& $U$ & 
   $({\bf 27},\2;\1)_{-1,0}$ +
   $(\1,\2;\1)_{3,0}$ +
   $(\1,\1;{\bf 64})_{0,\frac{1}{2}}$ +
2  $(\1,\1;\1)_{0,0}$
\\
 $
E_6 \times SU(2) \times U(1) 
\times SO(14) \times U(1)
$ 
& $T$ & 
$
\begin{array}{c}
12 (\1,\1;\1)_{\frac{3}{2},\frac{1}{2}} +
8  (\1,\2;\1)_{-\frac{3}{2},\frac{1}{2}} +\\
4  ({\bf 27},\1;\1)_{\frac{1}{2},-\frac{1}{2}} +
4  (\1,\1;{\bf 14})_{\frac{3}{2},-\frac{1}{2}}
\end{array}
$
\\
 $\frac{1}{4}(2,1^2,0^5;2,0^7)$ 
& $\!\!T^2\!\!$ &
3  $(\1,\2;{\bf 14})_{0,0}$ +
10 $(\1,\2;\1)_{0,1}$
\\\hline\hline 
IVf & $U$ & 
$\!\!\!\!\!\!({\bf 27},\2;\1,\1)_{-1}$ +
   $(\1,\2;\1,\1)_{3}$ +  
   $(\1,\1;{\bf 16},{\bf 4})_{0}$ +
2  $(\1,\1;\1,\1)_{0}\!\!\!\!\!\!$
\\
$
E_6 \times SU(2) \times U(1) 
\times SO(10) \times SO(6)
$ 
& $T$ & 
8  $(\1,\1;\1,{\bf 4})_{\frac{3}{2}}$ +
4  $(\1,\2;\1,{\bf 4})_{-\frac{3}{2}}$ +
4  $(\1,\1;{\bf\ov{16}},\1)_{\frac{3}{2}}$
\\
$\frac{1}{4}(2,1^2,0^5;2^3,0^5)$ 
& $\!\!T^2\!\!$ &
5  $(\1,\2;{\bf 10},\1)_{0}$ +
3  $(\1,\2;\1,{\bf 6})_{0}$
\\\hline\hline 
IVg & $U$ &
   $({\bf 12},\2;\1)_1$ + 
   $({\bf 32_+},\1;\1)_1$ +
2  $(\1,\1;\1)_0$
\\
 $SO(12) \times SU(2) \times U(1) \times E_8$ 
& $T$ & 
12 $(\1,\2;\1)_{-\frac{1}{2}}$ +
8  $({\bf 12},\1;\1)_{\frac{1}{2}}$ +
4  $(\1,\2;\1)_{\frac{3}{2}}$ +
4  $({\bf 32_-},\1;\1)_{-\frac{1}{2}}$
\\
 $\frac{1}{4}(3,1,0^6;0^8)$
& $\!\!T^2\!\!$ & 
32 $(\1,\1;\1)_{1}$ +
5  $({\bf 12},\2;\1)_{0}$ +
3  $({\bf 32_+},\1;\1)_{0}$
\\\hline\hline 
IVh & $U$ &
   $({\bf 12},\2;\1,\1)_1$ + 
   $({\bf 32_+},\1;\1,\1)_1$ +
2  $(\1,\1;\1,\1)_0$
\\
$SO(12) \times SU(2) \times U(1) \times E_7 \times SU(2)$
& $T$ & 
8 $(\1,\2;\1,\2)_{-\frac{1}{2}}$ +
4  $({\bf 12},\1;\1,\2)_{\frac{1}{2}}$
\\
$\frac{1}{4}(3,1,0^6;2^2,0^6)$
& $\!\!T^2\!\!$ & 
32 $(\1,\1;\1,\1)_{1}$ +
3  $({\bf 12},\2;\1,\1)_{0}$ +
5  $({\bf 32_+},\1;\1,\1)_{0}$
\\\hline\hline 
IVi & $U$ &
   $({\bf 12},\2;\1)_1$ + 
   $({\bf 32_+},\1;\1)_1$ +
2  $(\1,\1;\1)_0$
\\
 $SO(12) \times SU(2) \times U(1) \times SO(16)$
& $T$ & 
4 $(\1,\2;{\bf 16})_{-\frac{1}{2}}$ +
\\
$\frac{1}{4}(3,1,0^6;4,0^7)$
& $\!\!T^2\!\!$ & 
32 $(\1,\1;\1)_{1}$ +
5  $({\bf 12},\2;\1)_{0}$ +
3  $({\bf 32_+},\1;\1)_{0}$
\\\hline\hline 
IVj & $U$ & 
   $({\bf 28},\2;\1,\1)$ + 
   $(\1,\1;{\bf 16},\4)$ +
2  $(\1,\1;\1,\1)$
\\
$SU(8) \times SU(2) \times SO(10) \times SO(6) $
& $T$ & 
4  $({\bf 8};\1;\1,\4)$
\\
 $\frac{1}{4}(3,1^5,0^2;2^3,0^5)$ 
& $\!\!T^2\!\!$ &
5  $(\1,\2;\1,{\bf 6})$ +
3  $(\1,\2;{\bf 10},\1)$
\\\hline\hline
IVk & $U$ &
   $({\bf 28},\2;\1)_{0}$ + 
   $(\1,\1;{\bf 64})_{\frac{1}{2}}$+
2  $(\1,\1;\1)_{0}$
\\
$SU(8) \times SU(2) \times SO(14) \times U(1)$ 
& $T$ & 
8  $(\ov{\bf 8},\1;\1)_{\frac{1}{2}}$ +
4  $({\bf 8},\2;\1)_{\frac{1}{2}}$
\\
 $\frac{1}{4}(3,1^5,0^2;2,0^7)$
& $\!\!T^2\!\!$ &
5  $(\1,\2;{\bf 14})_0$ +
6  $(\1,\2;\1)_{1}$
\\\hline\hline 
IVl & $U$ & 
$\begin{array}{c}
   (\8;\1,\1)_{-3,0} + 
   ({\bf 56};\1,\1)_{-1,0} +
   (\1;{\bf 12},\2)_{0,1} +\\
   (\1;{\bf 32_+},\1)_{0,1} +
2  (\1;\1,\1)_{0,0}
\end{array}$
\\
$\!\!\!SU(8) \times U(1) \times SO(12) \times SU(2) \times U(1)\!\!\!$ 
& $T$ &
8  $(\1;\1,\2)_{2,-\frac{1}{2}}$ +
4  $(\1;{\bf 12},\1)_{2,\frac{1}{2}}$ +
4  $({\bf 8};\1,\2)_{-1,-\frac{1}{2}}$ +
\\
$\frac{1}{4}(1^7,-1;3,1,0^6)$ 
& $\!\!T^2\!\!$ & 
10 $(\8;\1,\1)_{1,1}$ +
6  $(\8;\1,\1)_{1,-1}$
\\\hline
\end{tabular}
\mbox{}\vspace{-.1cm}
\caption{
Perturbative $E_8 \times E_8$ spectra on $T^4/\Z_4$. The first column containing
shift vectors and gauge groups goes back to~\cite{Stieberger:1998yi}, the 
complete matter content is computed here for the first time. 
The standard embedding IVa is (up to charge normalization) taken
from~\cite{Erler:1993zy}. The spectrum of model IVj has been listed 
in~\cite{Kaplunovsky:1999ia} with the shift vector
$\frac{1}{4}(-7,1^7;-3,1^3,0^4)$, and IVf appears
in~\cite{Aldazabal:1995yw} with $\frac{1}{4}(1^2,-2,0^5;1^3,-3,0^4)$.}
\label{Tab:E8_Z4_Spectra}
\end{table}

\begin{table}[h!]
\renewcommand{\arraystretch}{1.2}
\begin{center}\footnotesize
\begin{tabular}{|c||l||c|}
\hline\hline
Shift \& gauge group && Matter\\
\hline\hline
6a & $U$ & 
   $({\bf 28},\2)_1$ + 
2  $(\1,\1)_0$\\
$SO(28)\times SU(2) \times U(1)$ & $T$ & 
8  $(\1,\1)_{\frac{1}{3}}$ +
2  $(\1,\1)_{-\frac{5}{3}}$ +
   $({\bf 28},\2)_{-\frac{2}{3}}$\\
$\frac{1}{6}\left(0^{14},1^2\right)$ & $T^2$ &
22 $(\1,\1)_{\frac{2}{3}}$ +
10 $(\1,\1)_{-\frac{4}{3}}$ +
5  $({\bf 28},\2)_{-\frac{1}{3}}$\\
& $T^3$ &
22 $(\1,\1)_1$ +
3  $({\bf 28},\2)_0$\\\hline
6b & $U$ & 
   $({\bf 12},\5,\1)_1$ + 
2  $(\1,\1,\1)_0$\\
$SO(12)\times SU(5)\times SO(10) \times U(1)$ & $T$ & 
2  $(\1,\1,{\bf 16})_{\frac{5}{6}}$\\
$\frac{1}{6}\left(0^{6},1^5,3^5\right)$ & $T^2$ &
$\left\{\begin{array}{c}
10 (\1,\5,\1)_{-\frac{2}{3}} +
4  (\1,\1,{\bf 10})_{\frac{5}{3}} +\\
5  ({\bf 12},\1,\1)_{\frac{5}{3}} +
4  (\1,{\bf 10},\1)_{-\frac{4}{3}}
\end{array}\right.$\\
& $T^3$ &
3  $({\bf 32_+},\1,\1)_0$\\\hline\hline
VIa & $U$ & 
   $({\bf 56},\1)_1$ + 
2  $(\1,\1)_0$\\
$E_7\times U(1)\times E_8$ & $T$ & 
8  $(\1,\1)_{\frac{1}{3}}$ +
2  $(\1,\1)_{-\frac{5}{3}}$ +
   $({\bf 56},\1)_{-\frac{2}{3}}$\\
$\frac{1}{6}\left(1^2,0^6;0^8\right)$ & $T^2$ &
22 $(\1,\1)_{\frac{2}{3}}$ +
10 $(\1,\1)_{-\frac{4}{3}}$ +
5  $({\bf 56},\1)_{-\frac{1}{3}}$\\
& $T^3$ &
22 $(\1,\1)_1$ +
3  $({\bf 56},\1)_0$\\\hline
VIb & $U$ & 
$\left\{\begin{array}{c}
   ({\bf 12},\1)_{1,0,0} + 
   ({\bf 32_-},\1)_{\half,-\half,0} +\\
   (\1,{\bf 64})_{0,0,\frac{1}{2}} +
2  (\1,\1)_{0,0,0}
\end{array}\right.$\\
$SO(12)\times SO(14)\times U(1)^3$ & $T$ & 
$\left\{\begin{array}{c}
5  (\1,\1)_{\frac{1}{6},\frac{1}{2},\frac{1}{3}} +
3  (\1,\1)_{-\frac{5}{6},-\frac{1}{2},\frac{1}{3}} +\\
   (\1,\1)_{\frac{7}{6},-\frac{1}{2},\frac{1}{3}} +
2  ({\bf 12},\1)_{\frac{1}{6},-\frac{1}{2},\frac{1}{3}} +\\
   (\1,{\bf 14})_{\frac{1}{6},\frac{1}{2},-\frac{2}{3}} +
   ({\bf 32_+},\1)_{-\frac{1}{3},0,\frac{1}{3}} +
\end{array}\right.$\\
$\frac{1}{6}\left(0^6,1,3;0^7,2\right)$ & $T^2$ &
$\left\{\begin{array}{c}
10  (\1,\1)_{-\frac{2}{3},0,\frac{2}{3}} +
5  ({\bf 12},\1)_{\frac{1}{3},0,\frac{2}{3}} +\\
4   (\1,{\bf 14})_{-\frac{2}{3},0,-\frac{1}{3}} +
4  (\1,\1)_{\frac{1}{3},\pm 1,\frac{2}{3}}
\end{array}\right.$\\
& $T^3$ &
 $\left\{\begin{array}{c}5 (\1,\1)_{\frac{1}{2},-\frac{1}{2},1} 
 +6  (\1,\1)_{\frac{1}{2},-\frac{1}{2},-1}\\
 +5  (\1,{\bf 14})_{\frac{1}{2},-\frac{1}{2},0}\end{array}\right.$
\\\hline
VIc & $U$ & $(\6,\3,\2;\1) + 2(\1)$
\\
$SU(6) \times SU(3) \times SU(2) \times SU(9)$ & $T$ & 
$(\6,\1,\1;\ov{\9})$
\\
$\frac{1}{6}(-5,1^5,0^2;-5,1^7)$ & $T^2$ &
$4(\1,\3,\1;\9)$
\\
& $T^3$ &
$3({\bf 20},\1,\1;\1)+ 5 (\6,\3,\1;\1)+ 10(\1,\1,\2;\1)$
\\\hline
VId & $U$ & $\left\{\begin{array}{c}(\6,\3,\2;\1,\1)_0 + 2(\1)_0 +\\ 
(\1,\1,\1;{\bf 10},\4)_1 + (\1,\1,\1;\1,\4)_5\end{array}\right.$
\\
$\begin{array}{c} 
SU(6) \times SU(3) \times SU(2) \\ 
\times SU(5) \times SU(4) \times U(1)\end{array}$  
& $T$ & 
$\begin{array}{c}
2(\6,\1,\1;\1,\1)_{\frac{10}{3}} 
+ (\6,\1,\1;\1,\ov{\4})_{-\frac{5}{3}} + 
(\1,\3,\2;\1,\1)_{\frac{10}{3}}
\end{array}$
\\
$\frac{1}{6}(-5,1^5,0^2;-4,1^4,0^3)$ 
& $T^2$ & 
$4(\1,\3,\1;\5,\1)_{\frac{4}{3}} + 5 (\1,\3,\1;\1,\4)_{-\frac{5}{3}}$
\\
& $T^3$ & $5(\1,\1,\2;\5,\1)_{-2}+3(\1,\1,\2;\1,\6)_0$
\\\hline
\end{tabular}
\end{center}
\caption{Gauge group and matter for two $SO(32)$ and four
$E_8\times E_8$  heterotic string models on $T^4/\Z_6$. The standard
embedding VIa can be found in~\cite{Erler:1993zy} (up to charge normalization), model VIc is taken 
from~\cite{Kaplunovsky:1999ia} and VId from~\cite{Aldazabal:1995yw}.}
\label{Tab:Z6both}
\end{table}

\paragraph{Twisted matter\\}
Neglecting the gauge enhancement $SO(16)\rightarrow E_8$,
the computation of the twisted matter can be done exactly along
the lines explained in section~\ref{Sec:SO}  for the $SO(32)$ case.
We are then able to write tables similar to \ref{Tab:SO(32)_Z4a}
and \ref{Tab:SO(32)_Z4b} (\ref{Tab:SO(32)_Z4red} in the $g^2$-twisted
case) that summarize in an abstract way the whole set of models.
Of course, in this approach the twisted matter is arranged in representations
of the subgroup of $SO(16)\times SO(16)$ that survives the orbifold
projection. If the latter is enhanced, we  have to recombine the
twisted states into representations of the enhanced gauge
symmetry as well.

The complete massless spectra are listed in table~\ref{Tab:E8_Z4_Spectra}.

\subsection{Examples of $T^4/\Z_6$ orbifold vacua}

The 60 inequivalent shift vectors for $E_8 \times E_8$ orbifolds on
$T^4/\Z_6$ have been classified in~\cite{Stieberger:1998yi}, for the
gauge group $SO(32)$ roughly the same number of different spectra is
expected.
The computation of gauge symmetry breaking and untwisted spectrum
is exactly as described in sections~\ref{Sec:SO}, with the caveat that,
in the $E_8\times E_8$ case, a gauge enhancement is generically
expected, as shown previously in the $\Z_4$ case in table~\ref{Tab:E8_Z4_shifts}.
About the twisted spectrum, it contains three sectors, that can be
computed following the approach used in the $\Z_4$ case. In detail,
the only new sector is the $g$-twisted one, since the $g^2$-twisted
sector corresponds to a $\Z_6$ projection of the $g$-twisted sector
of the $\Z_3$ model, and the $g^3$-sector in a projection of the
twisted sector of the $\Z_2$ model.
We do not attempt  a complete classification of these models, but rather
give a few examples, summarized in table~\ref{Tab:Z6both}.

\subsection{Anomaly polynomials for the orbifold models}
\label{Sec:Ano}

The anomaly polynomial is computed for each spectrum separately along the lines described in detail in~\cite{Erler:1993zy} up to the overall
normalization change of the anomaly polynomial by a factor of $(-16)$ which leads to a prefactor 1 of the gravitational part in 
the expansion for perturbative vacua \mbox{$I_8 = (\tr R^2)^2 + \ldots$}. This normalization agrees with the one used 
in~\cite{Honecker:2006dt} for 
the smooth $K3$ embeddings with which the orbifold point anomaly polynomials will be compared in section~\ref{Sec:Bundles}.

It turns out that there occurs a very general form, namely for $SO(32)$ heterotic orbifold models we obtain
\bea
I_8^{SO(32)} &\hspace{-4pt}=& \hspace{-4pt}
\left( \tr R^2 + \sum_i \alpha_i \tr_{SO(2M_i)} F^2 + \sum_j \beta_j \tr_{SU(N_j)} F^2  
+\sum_k \gamma_k F_{U(1)_k}^2
+ \sum_{i<j} \delta_{ij} \,  F_{U(1)_i} \, F_{U(1)_j} 
\right) \times 
\nonumber\\
& &\hspace{-8pt}
\times \left(\tr R^2 - \sum_i  \tr_{SO(2M_i)} F^2  -2 \, \sum_j  \tr_{SU(N_j)} F^2 
+\sum_k \tilde{\gamma}_k F_{U(1)_k}^2
\right)
\label{Eq_Anomaly_Orbifold_SO32}
\eea
with the coefficients $\alpha,\beta,\gamma, \delta$ and $\tilde{\gamma}$ listed in tables~\ref{Tab:SO32factorisation_coeffs_normal},
~\ref{Tab:SO32factorisation_coeffs_exc_1} and~\ref{Tab:SO32factorisation_coeffs_exc_3}.

\begin{table}[h]
\renewcommand{\arraystretch}{1.2}
\begin{center}\footnotesize
\begin{tabular}{|c||c|c||c|c||c|c|}
\hline\hline
\# & $2M$ & $\alpha$ & $N$ & $\beta$ & $\gamma$ & $\tilde{\gamma}$ 
\\\hline\hline
2c & - & - & 16 & 0 & -512 & -32  
\\\hline\hline
3,4,6a & 28 & 2  & 2 & -44 & -24 & -4 
\\\hline
3b & 20 & 2  & 5 & -8 & -140 & -10
\\\hline
3c & 16 & 2  & 8 & -8  & -64 & -16 
\\\hline
3d & 10$^{\star}$ & $-\frac{5}{2}$  & 11 & 1 & -231 & -22
\\\hline\hline
4a' & 28 & 2  & 2 & -28 & -24 & -4 
\\\hline
4e & 12$^{\star}$  & -1  & 10 & -2 & -20 & -20 
\\\hline
4e' & 12$^{\star}$  & -3  & 10 & 2 & -80 & -20
\\\hline
\end{tabular}
\end{center}
\caption{Coefficients in the $SO(32)$ anomaly polynomials
for models with up to one gauge group of each kind, $SO(2M) \times SU(N) \times U(1)$.
A star at the value of $2M$  indicates that the massless spectrum contains spinor representations of $SO(2M)$.
For all models, we have $\tilde{\gamma} = -2N$, which coincides with the expected value for the $U(1)$ to be 
the trace part of an $U(N)$ gauge factor. However, $\gamma$ does not match with the smooth models discussed in section~\ref{Subsec:SO32bundles},
and from the explicit spectra in tables~\ref{Tab:SO32_Pert_Orbifold_Spectra},~\ref{Tab:SO(32)_Z4allinall},~\ref{Tab:SO(32)_Z4spiallinall} 
and~\ref{Tab:Z6both}, one sees  that instead of being the trace part of some larger group, the $U(1)$ charges are associated to the 
twist sectors of the orbifold.
}
\label{Tab:SO32factorisation_coeffs_normal}
\end{table}

\begin{table}[h]
\renewcommand{\arraystretch}{1.2}
\begin{center}
\begin{tabular}{|c||c|c||c|c||c|c||c|c||c|c||c|c|}
\hline\hline
\# & $2M_1$ & $\alpha_1$
  & $2M_2$ & $\alpha_2$ 
& $N_1$ & $\beta_1$  
& $N_2$ & $\beta_2$
& $N_3$ & $\beta_3$  
& $\gamma$ & $\tilde{\gamma}$ 
\\\hline\hline
2a & 28 & 2 &  - & - & 2 & -44 & 2 & -12  & - & - & - & - 
\\\hline
2b & 20 & 2 & 12$^{\star}$  & -6  & - & - & - & - & - & - & - & - 
\\\hline\hline
3e & - & - & - & - &  14 & 1 & 2 & -23 &  2 & -5  & -182 & -28 
\\\hline\hline
4b & 24 & 2 & - & -  &  2 & -20 & 2 & -4 &  2 & -20  & -24 & -4 
\\\hline
4c & 20 & 2 &  8$^{\star}$   & -2 & 2 & -28 & - & -& - & -  & -16 & -4  
\\\hline
4d & 16 & 2 &  12$^{\star}$  & -2 & 2 & -20  & - & - & - & - & -16 & -4 
\\\hline
4f & 8$^{\star}$  & -3 & - & - &  10 & 2 & 2 & -6 &  2 & -14 & -80 & -20 
\\\hline
4h & 14 & 2 & - & - & 6$^{\star}$ & -8 & 4 & -4& - & -  & -48 & -12 
\\\hline
4i & 10$^{\star}$  & 0  & 10$^{\star}$  & 0 & 6$^{\star}$ & -4 & - & - & - & - & -96 & -12 
\\\hline\hline
6b & 12$^{\star}$  & -1 & 10$^{\star}$  & 1 & 5  & -2 & - & -& - & -  & $-\frac{590}{9}$ & -10 
\\\hline
\end{tabular}
\end{center}
\caption{Coefficients for models with several $SO(2M)$ and $SU(N)$
factors. For all models with an $U(1)$ gauge factor, we have
$\tilde{\gamma} = -2N_1$. A star indicates that the massless spectrum
contains representation of the gauge groups that cannot be reproduced
in the smooth case (spinorial representation of $SO(2M)$, third
rank totally antisymmetric representation of $SU(N)$ etc.).
}
\label{Tab:SO32factorisation_coeffs_exc_1}
\end{table}


\begin{table}[h!]
\renewcommand{\arraystretch}{1.2}
\begin{center}
\begin{tabular}{|c||c|c||c|c||c|c||c|c||c|c||c|c||c|c|c|}
\hline\hline
\# & $2M$ & $\alpha$ 
& $N_1$ & $\beta_1$   
& $N_2$ & $\beta_2$  
& $\gamma_1$ & $\tilde{\gamma}_1$
& $\gamma_2$ & $\tilde{\gamma}_2$ 
& $\gamma_3$ & $\tilde{\gamma}_3$ 
& $\delta_{12}$  
& $\delta_{13}$ & $\delta_{23}$   
\\\hline\hline
4g & 18 & 2  & 6$^{\star}$ & -8  & - & - &  -84 & -12 & -8 & -2 & - & - & 24 & - & - 
\\\hline
4j & - & - & 14 & 2  & - & -& -16 & -2 & -168 & -28 & -16 & -2 & 28 & -16 & 28 
\\\hline
4k & - & -  & 15 & 1  & - & -  & $-\frac{61}{8}$ & -2 & $-\frac{2805}{8}$ & -30 & - & -  & $\frac{15}{4}$ & - & - 
\\\hline
4l & - & -  & 11 & 1  & 5 & -3  & $-\frac{595}{8}$ & -10 & $-\frac{1243}{8}$ & -22 & - & -  & $\frac{385}{4}$ & - & - 
\\\hline
4m & - & - & 9 & 1  & 7 & -3  & $-\frac{1143}{8}$ & -18 & $-\frac{511}{8}$ & -14 & - & -  & $\frac{189}{4}$ & - & - 
\\\hline
4n & - & -  & 13 & 1  & 3 & -11  & $-\frac{1417}{8}$ & -26 & $-\frac{129}{8}$ & -6 & - & -  & $\frac{195}{4}$ & - & - 
\\\hline
\end{tabular}
\end{center}
\caption{Coefficients for models with several $U(1)$ factors. For 4g, 4j, 4m, 4n we have $\tilde{\gamma}_1=-2N_1$, for 4k, 4l 
$\tilde{\gamma}_2=-2N_1$. Furthermore, for 4l we have $\tilde{\gamma}_1=-2N_2$, and for 4m, 4n $\tilde{\gamma}_2=-2N_2$. }
\label{Tab:SO32factorisation_coeffs_exc_3}
\end{table}

Also for the $E_8 \times E_8$ case, the anomaly polynomial can be cast into a generic form:
label non-Abelian groups of rank $r_i$ descending from $E_8^{(i)}$ ($i=1,2$) by $G_{r_i}$.
Then the anomaly polynomial for $E_8 \times E_8$ orbifold compactifications to six dimensions has the general form
\bea
I_8^{E_8 \times E_8} 
&=&  \left( \tr R^2 - \sum_{i=1}^2 \sum_{x} a_{r_i}^x \tr_{G_{r_i}^x} F^2 - \sum_{y} b_y F_{U(1)_y}^2 
+ \sum_{y < z} c_{yz} F_{U(1)_y} \,F_{U(1)_z} 
\right) \times 
\nonumber\\
&&
\times \left(\tr R^2 -\sum_{i=1}^2 \sum_{x} \tilde{a}_{r_i}^x \tr_{G_{r_i}^x} F^2 - \sum_{y} \tilde{b}_y F_{U(1)_y}^2 
\right).
\label{E8:Anomaly_Polynomial_General}
\eea
The coefficients $a_{r_i}^x,  b_y, c_{yz},\tilde{b}_y $ 
depend on the combination of two  shift vectors and are listed in table~\ref{Tab:E8factorisation_systematics}, whereas the 
coefficients $\tilde{a}^x_r$ are universal for fixed gauge groups,
\bea
\begin{array}{|c||c|c|c|c|c|}
\hline
G_r & E_8 & E_7 & E_6 & SO(2M) & SU(N)
\\\hline
\tilde{a}_r & 1 & \frac{1}{6} & \frac{1}{3} & 1 & 2
\\\hline
\end{array}
\label{Eq:coeff_E8_atilde}
\eea
and coincide for $SO(2M)$ and $SU(N)$ with those of the $SO(32)$ heterotic orbifolds.
It turns out that the instanton number $k$ inside an $E_8$ gauge factor
(with $k_1 + k_2 = 24$) listed in table~\ref{Tab:E8factorisation_systematics} is
for all $T^4/\Z_N$ models with $N=2,3,4$ related to the coefficients of the largest
non-Abelian gauge factor by
\bea
k= 12 + 2 \, \frac{a^1_r}{\tilde{a}^1_r}.
\eea
Assuming that this relation holds also for $N=6$, the coefficient $a^1_r$ of the largest
gauge factor inside each $E_8$ can be computed for all models from the instanton numbers
given in~\cite{Stieberger:1998yi} without having to compute the matter spectrum.

For both gauge groups $SO(32)$ and $E_8 \times E_8$, the 
anomaly polynomials~(\ref{Eq_Anomaly_Orbifold_SO32}) and~(\ref {E8:Anomaly_Polynomial_General})
factorize completely into $4 \times 4$ forms. This is in contrast to smooth compactifications where a sum of 
two factorized expressions $ 4 \times 4 + 6 \times 2$ occurs, and the second part signals 
that $U(1)$ gauge factors become massive via the Green-Schwarz mechanism.
It is therefore natural to deduce that the $U(1)$ gauge groups at the orbifold point remain massless, however, it should be stressed that 
the absence of a $6 \times 2$ factorized part could in principle also be due to the absence of the six-form part in the presence of a mass term of the
$U(1)$ gauge fields providing the two-form part.

Apart from the discrepancy of the Abelian gauge factors, we will see a very similar pattern for the anomaly polynomials in the smooth case
in section~\ref{Sec:Bundles}, and the matching of coefficients will be a guiding principle for obtaining the correct second Chern characters
(instanton numbers) of the bundles.
\clearpage
\begin{table}[h!]
\renewcommand{\arraystretch}{1.2}
\footnotesize\mbox{}\vspace{-.1cm}\mbox{}\hspace{-.7cm}
\begin{tabular}{|c||c|c|c||c|c||c|c||c||c|}
\hline\hline
 & $G_r^1$ & $a_r^1$ & $\!\!\!\!\ch_2(L)\!\!\!\!$ & $G_r^2$ & $a_r^2$  & $b_y^1$ & $ \tilde{b}_y^1$
& $c_{12}$ & $k$
\\\hline\hline
$\begin{array}{c}
(0^8)\\
\mbox{II-VIa, IIIc, IVg} 
\end{array}$
& $E_8$ & -6 & 0 & - & - & - & - & - & $0$
\\\hline\hline
$\begin{array}{c}
\frac{1}{N}(1^2,0^6) \\ N\neq 2
\end{array}
\begin{array}{c}
\mbox{III-VIa}\\
\mbox{IIId, IVd} \\
\mbox{IVb}\\
\mbox{IVc}
\end{array}$
& $E_7$ & $\begin{array}{c} 1\\ -\half \\ 0 \\ \frac{1}{3} \end{array}$ 
&  $\begin{array}{c} -12 \\ -3 \\ -6 \\ -8 \end{array}$
& - & - 
& $\begin{array}{c} 24 \\ 36 \\ 24 \\ 16   \end{array}$  & 4 & -
& $\begin{array}{c} 24 \\ 6 \\ 12 \\ 16 \end{array}$
\\\hline
$\frac{1}{2}(1^2,0^6)\quad
\begin{array}{c}
\mbox{IIa}\\
 \mbox{IIb}\\
\mbox{IVb}\\
\mbox{IVh}
\end{array}$
& $E_7$ & $\begin{array}{c} 1\\-\frac{1}{3} \\ 0 \\ -\frac{2}{3}\end{array}$ 
& $\begin{array}{c} -12 \\ -4 \\ -6 \\ -2 \end{array}$
& $SU(2)$ 
& $\begin{array}{c} 12 \\ 28^{\star} \\ 0 \\ 8^{\star} \end{array}$ & - & - & -
& $\begin{array}{c} 24 \\ 8 \\ 12 \\ 4  \end{array}$
\\\hline\hline
$\frac{1}{3}(1^2,2,0^5) \quad 
\begin{array}{c}
\mbox{IIId}\\
\mbox{IIIe}
\end{array}$ 
& $E_6$ & $\begin{array}{c} 1\\ -\half \end{array}$  
& $\begin{array}{c} -3 \\ -\frac{3}{2} \end{array}$ 
& $SU(3)$ & $\begin{array}{c} 6\\15^{\star} \end{array}$   
 & - & - & - 
& $\begin{array}{c} 18 \\ 9 \end{array}$   
\\\hline
$\frac{1}{4}(1^2,2,0^5) \quad 
\begin{array}{c}
\mbox{IVe}\\
\mbox{IVf}
\end{array}$ 
& $E_6$ & $\begin{array}{c} 0\\-\frac{2}{3} \end{array}$  
&  $\begin{array}{c} -2 \\ - \frac{4}{3} \end{array}$   
& $SU(2)$ 
&  $\begin{array}{c} 12^{\star} \\ 16^{\star}  \end{array}$    & $\begin{array}{c} 36\\ 48 \end{array}$ & 12 
&  $\begin{array}{c} 24 \\ -  \end{array}$ 
&  $\begin{array}{c} 12 \\ 8  \end{array}$ 
\\\hline\hline
$\begin{array}{c}
\frac{1}{2}(1,0^7)\quad \mbox{IIb}\\
(1,0^7)\quad \mbox{IVc,i}
\end{array}$
& $SO(16)$ & $\!\!\!\!\begin{array}{c} 2^{\star}\\ -2^{(i\star)}  \end{array}\!\!\!\!$  
&  $\begin{array}{c} -4 \\ -2  \end{array}$   
 & - & - & - & - & -
&  $\begin{array}{c} 16 \\ 8  \end{array}$ 
\\\hline
$\begin{array}{c}
\frac{1}{N}(2,0^7) \\
N \neq 2 
\end{array}
\begin{array}{c}
\mbox{IIIb}\\
\mbox{IVe}\\
\mbox{IVk}\\
\mbox{VIb}
\end{array}$ 
& $SO(14)$ & $\begin{array}{c} 0 \\ 0^{\star} \\  0^{\star} \\  0 \end{array}$
& -3
& - & -  & $\begin{array}{c} 16 \\ 12 \\ 8 \\ \frac{104}{9}\end{array} $ & 2 
& $\begin{array}{c} 16 \\ 24 \\ - \\  -\frac{56}{9}, \frac{8}{3}\end{array}$
& $12$
\\\hline\hline
$\frac{1}{4}(3,1,0^6) \quad
\begin{array}{c}
\mbox{IVg}\\
\mbox{IVh}\\
\mbox{IVi}\\
\mbox{IVl}
\end{array}$
& $SO(12)$ & $\begin{array}{c} 6\\4^{\star} \\ 2 \\-1 \end{array}$ &  
$\begin{array}{c} -\frac{12}{5} \\ -2 \\ -\frac{8}{5} \\ -1  \end{array}$
& $SU(2)$ & $\begin{array}{c} 12\\8^{\star}  \\ 20^{\star}  \\ 6^{\star} \end{array}$ 
& $\begin{array}{c} 24 \\ 16 \\ 16 \\ 32\end{array}$   & 4  & 
 $\begin{array}{c} - \\ - \\ - \\ -32\end{array}$  
& $\begin{array}{c} 24 \\ 20 \\ 16 \\ 10 \end{array}$
\\\hline
$\frac{1}{6}(3,1,0^6) \quad
\mbox{VIb}$
& $SO(12)$ & 0 &   $-\frac{6}{5}$
& - & - &   $\begin{array}{c} \frac{110}{9} \\ 6 \end{array}$ & 2  
&  $\!\!\frac{28}{3}, \!\!\begin{array}{c} \frac{-56}{9} \\ \frac{8}{3} \end{array}\!\!\!$ 
& 12
\\\hline\hline
$\frac{1}{4}(2^3,0^5) \quad
\begin{array}{c}
\mbox{IVf}\\
\mbox{IVj}
\end{array}$
& $SO(10)$ & $\begin{array}{c} 2^{\star}\\0^{\star}\end{array}$  &
 $\begin{array}{c} -\frac{4}{3} \\ -1 \end{array}$ 
 & $SU(4)$ & $\begin{array}{c} 4^{\star}\\8^{\star}\end{array}$  
 & - & - & - 
& $\begin{array}{c} 16 \\ 12  \end{array}$
\\\hline
$\frac{1}{3}(1^4,2,0^3) \quad 
\begin{array}{c}
\mbox{IIIc}\\
\mbox{IIIe}
\end{array}$ 
& $SU(9)$ & $\begin{array}{c} 12\\3^{\star} \end{array}$ &   
 $\begin{array}{c} -3 \\ -\frac{15}{8}  \end{array}$
& - & - & - & - & -
& $\begin{array}{c} 24 \\ 15 \end{array}$
\\\hline
$\frac{1}{6}(-5,1^7)\quad 
\mbox{VIc}$
& $SU(9)$ & $-2^{\ast}$ & $-\frac{5}{16}$  & - & - & - & - & - 
& 10$$
\\\hline
$\frac{1}{4}(1^7,-1) \quad 
\begin{array}{c}
\mbox{IVd}\\
\mbox{IVl}
\end{array}$ 
& $SU(8)$ &  $\begin{array}{c} 6\\ 2^{\star} \end{array}$  &  
 $\begin{array}{c} -\frac{9}{4} \\ -\frac{7}{4}  \end{array}$  
&  - & - &  $\begin{array}{c} 48\\ 80 \end{array}$ 
& 16 & $\begin{array}{c} - \\ -32 \end{array}$
&  $\begin{array}{c} 18 \\ 14 \end{array}$
\\\hline
$\frac{1}{4}(3,1^5,0^2) \quad
\begin{array}{c}
\mbox{IVj}\\
\mbox{IVk}
\end{array}$
& $SU(8)$ &  $\begin{array}{c}0^{\star}\\ 0\end{array}$ &  $-\frac{6}{7}$
& $SU(2)$ 
& $\begin{array}{c} 12^{\star} \\ 20^{\star}\end{array}$  & - & - & - 
&  12$$ 
\\\hline
$\frac{1}{6}(-5,1^5,0^2)  \quad
\begin{array}{c}
\mbox{VIc}\\
\mbox{VId}
\end{array}$
& $SU(6)$ & $\begin{array}{c} 4^{\star} \\ -2^{\star} \end{array}$ &  $\begin{array}{c} -\frac{7}{15}\\ -\frac{1}{3} \end{array}$
& $\begin{array}{c} SU(3) \\ SU(2) \end{array}$ &
$\!\!\!\!\!\begin{array}{c} 
{\rm VIc:}\!\left\{\begin{array}{c} \!10^{\star}\\ \!2
\end{array}\right.
\\
{\rm VId:}\!\left\{\begin{array}{c}
\!6^{\star} \\ \!8^{\star}
\end{array}\right.
  \end{array}\!\!\!\!\!\!\!\!$
& - & - & -
&  $\begin{array}{c} 14 \\ 10 \end{array}$
\\\hline
$\frac{1}{6}(-4,1^4,0^2)  \quad
\mbox{VId}$
& $SU(5)$ & $2^{\star}$ & $-\frac{7}{10}$ & $SU(4)$ & $4^{\star}$ & $\frac{320}{3}$ & 40 & - & 14
\\\hline
\end{tabular}
\mbox{}\vspace{-.1cm}
\caption{Systematics on coefficients for $E_8$. A star at the value of $a_r$
indicates that there exists matter charged under both this gauge group
as well as some gauge factor which arises from the other $E_8$
breaking; $k$  labels the instanton numbers from one $E_8$ factor given in~\cite{Aldazabal:1997wi,Stieberger:1998yi,Kaplunovsky:1999ia}.
The entries $\ch_2(L)$ correspond to smooth line bundle matchings and are explained in section~\ref{Subsec:E8_K3_U1s}.
}
\label{Tab:E8factorisation_systematics}
\end{table}\thispagestyle{empty}
\clearpage

\section{The heterotic string on $K3$ with line bundles}
\label{Sec:Bundles}

In this chapter, the six dimensional heterotic orbifold spectra constructed in section~\ref{Sec:Orbifolds} are compared 
with smooth compactifications on $K3$ with line bundles. After a general discussion of 
consistency conditions in section~\ref{Subsec:K3general}, six dimensional model building on $K3$ of
the $SO(32)$ heterotic string is elaborated on in section~\ref{Subsec:SO32bundles}
and  for the $E_8 \times E_8$ heterotic string in section~\ref{Subsec:E8_K3_U1s}.
Some explicit matchings of spectra with the orbifold cases are presented for both classes,
and it is argued that for other cases there exist obstructions to find smooth matchings with the simple ansatz 
of embedding just one line bundle presented here. We infer that orbifold models with several twist sectors 
require multiple line bundles.

\subsection{6D spectra, supersymmetry and tadpole cancellation}
\label{Subsec:K3general}

We review briefly the basic model building features for both $SO(32)$ and $E_8 \times E_8$ 
heterotic string compactifications on $K3$ discussed in detail in~\cite{Honecker:2006dt}:
\begin{itemize}
\item Each hyper multiplet transforming in some representation ${\bf R}$ is associated to a bundle $V$ on $K3$. (Minus) the number of 
hyper multiplets is given by the Riemann-Roch-Hirzebruch theorem,
\bea
\chi(V)_{K3} \equiv \int_{K3} \ch (V) \, {\rm Td } (K3) 
=  \ch_2(V) + 2 r ,
\label{RRH}
\eea
where $\ch (V)=r+c_1(V) +\ch_2(V)+\ldots$ is the total Chern character of the bundle $V$, $r$ its rank and 
${\rm Td}(K3)=1+\frac{1}{12} c_2(K3) + \ldots = 1 + 2 \, {\rm vol}_{4} + \ldots$ 
the Todd class of the tangent bundle of the $K3$ surface.\footnote{Whenever $\ch_2(V)$ appears in an index, integration over $K3$ is understood 
since any four form is proportional to the normalized volume form, $\int_{K3} {\rm vol}_4=1$.}
Since the sign of the supersymmetric index~(\ref{RRH}) corresponds to the chirality of the fermions in a multiplet, in this convention, a {\it negative} index counts the number 
of {\it hyper} multiplets, while a {\it positive} index implies the existence of {\it vector} multiplets in the representation ${\bf R}$. \\
The assignment of representations and bundles will be discussed in section~\ref{Subsec:SO32bundles} for $SO(32)$ embeddings 
and~\ref{Subsec:E8_K3_U1s} for $E_8 \times E_8$ cases.
\item
Associated to the bundle $V$ is a background field strength $\ov{F}$ on $K3$. A supersymmetric background requires
\begin{enumerate}
\item
$\ov{F}$ is holomorphic, i.e. a (1,1) form and has no contributions from the (2,0) or (0,2) form.
\item
$\ov{F}$ is primitive, 
\bea
\int_{K3} J \wedge \tr \ov{F} =0,
\label{eq:primitive}
\eea
where $J$ is the K\"ahler form on $K3$. The $K3$ lattice is of type (3,19), which means that there 
exist three self dual and 19 anti-selfdual forms. For the K\"ahler form $J$ to be well defined, it 
has to lie in the self dual part of the lattice, and (\ref{eq:primitive}) is satisfied if $\tr \ov{F}$ 
is zero or $\ov{F}$ lies in the anti-self dual sublattice.
\end{enumerate}
\item
The supersymmetry conditions are trivially satisfied for $SU(n)$ bundles which have vanishing first Chern class.
For more general $U(n)$ bundles, the holomorphicity condition freezes two geometric moduli, and the primitivity 
fixes a third modulus.
In order to preserve supersymmetry,
the  $U(1)$ factor inside an observable $U(N)$ must become massive by absorbing a complete neutral hyper multiplet. 
The fourth scalar d.o.f. inside such a  neutral hyper multiplet is given by the dimensional reduction 
$b^{(0)}_k$ of the ten dimensional 
antisymmetric tensor $B$ over a two-cycle inside $K3$ labeled by the index $k$. The tree level couplings
\bea
S_{\rm mass} =  \sum_{k=0}^{21} \,  \frac{1}{4 \pi \ell_s^4} 
\int_{\mathbb{R}^{1,5}} c_{k}^{(4)} \wedge \left[ \tr (F \ov{F}) \right]^{(k)}
\label{Eq:U1masses}
\eea
between the four forms $c_{k}^{(4)}$ dual to the scalars $b^{(0)}_k$ 
and the $U(1)$ gauge field generate the mass.\\
Since $\tr (F \ov{F}) = \sum_i a_i c_1(V_i) \; F_{U(1)_i}$ holds with the coefficients $a_i$ depending on the specific embedding, 
in the generic case of different $V_i$, all $U(1)$ gauge factors become massive, and linear combinations of the $U(1)$s 
remain massless only if the first Chern classes 
of the respective bundles are linearly dependent.
\item
The Bianchi identity on the three form field strength $H= dB - \frac{\alpha'}{4} (\omega_{YM} - \omega_{L})$
results in the so called ``tadpole cancellation condition'' on the background fields,
\bea
\tr \ov{F}^2 - \tr \ov{R}^2 = 0
\label{tcc}
\eea
in cohomology. 
In this article, we restrict ourselves to {\it perturbative} vacua only; the generalization to including H5-branes is
straight forward as discussed in~\cite{Honecker:2006dt}.\\
The Bianchi identity is replaced in orbifold compactifications by the quadratic level matching condition
on the shift vectors~(\ref{Eq:Orbi_Level}).
\item
The so called ``K-theory constraint'' requires
\bea
c_1(W_{total}) \in H^2(K3,2\Z)
\eea 
for the total bundle $W_{total}$ to admit spinors.\\
In the orbifold case, this corresponds to the linear modular invariance constraint on the shift vectors~(\ref{Eq:Orbi_Modular}).
\item
The Green-Schwarz mechanism consists of two types of counter diagrams:
\begin{figure}
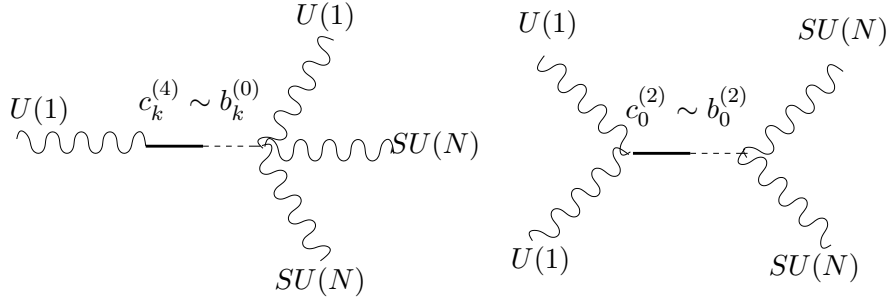

\begin{center}
\input GS_6D.pstex_t
\end{center}
\caption{The two possible types of Green-Schwarz counter diagrams. 
The couplings on the left in the $2 \times 6$ factorized diagram generate  $U(1)$ masses.}
\label{Fig:GS}
\end{figure}
the eight-form anomaly polynomial is a sum of a factorization as $2 \times 6$ and another as $4 \times 4$ forms,
\bea
{\cal I}_{pert} &=& \frac{1}{48 (2 \pi \ell_s)^4}\int_{K3}\left(
\tr (F \ov F) \wedge X_{\ov{2}+6} + \frac{1}{2}\left(\tr F^2 - \tr R^2 \right)\wedge  X_{\ov{4}+4}
\right),
\label{E_GS_pert}
\eea
respectively, as shown in figure~\ref{Fig:GS}. The first term in~(\ref{E_GS_pert}) corresponds to the exchange of the four forms $c^{(4)}_k$
and their scalar duals $b^{(0)}_k$ and involves always at least one Abelian gauge factor.
The last term has the two form $c^{(2)}_0$ and its dual $b^{(2)}_0$ (which is $B$ truncated to six dimensions) 
as internal propagating fields and contributes to the cancellation of pure and mixed gravitational and non-Abelian anomalies 
as well as Abelian ones. $X_{\ov{m}+n}$ labels the eight form appearing in the ten dimensional Green Schwarz counter 
term with $n$ indices along $\mathbb{R}^{1,5}$ and $\ov{m}$ indices on $K3$.
\end{itemize}

\subsection{$U(n)$ bundles inside $SO(32)$}
\label{Subsec:SO32bundles}

The starting point for matching $SO(32)$ heterotic orbifold vacua with smooth $K3$ compactifications is 
the group theoretical decomposition
\mbox{$SO(32) \rightarrow  SO(2M) \times \prod_i U(N_i \,n_i)$} and its adjoint 
representation,\footnote{For the same kind of decomposition in the context of $CY_3$-fold compactifications see~\cite{Blumenhagen:2005pm}.}
\bea
{\bf 496} \to
\left(\begin{array}{c} 
(\Anti_{SO(2M)})\\
\sum_{j=1}^{K} (\Adj_{U(N_j)};\Adj_{U(n_j)})\\
\sum_{j=1}^{K} (\Anti_{U(N_j)};\Sym_{U(n_j)}) + 
               (\Sym_{U(N_j)};\Anti_{U(n_j)}) + c.c.\\
\sum_{i < j} (\N_i,\N_j;\n_i,\n_j) + (\N_i,\ov{\N}_j,\n_i,\ov{\n}_j) + c.c. \\
\sum_{j=1}^{K} ({\bf 2M}, \N_j;\n_j) + c.c.\\
\end{array}\right),  
\eea
where $M \equiv 16 -\sum_i N_i \,n_i$.
Embedding $U(n_i)$ bundles $V_i$ inside $U(N_i \, n_i)$ leads to the massless spectrum listed in table~\ref{TchiralSO32},
\begin{table}[htb]
\renewcommand{\arraystretch}{1.5}
\begin{center}
\begin{tabular}{|c||c|}
\hline
\hline
reps. & $H= SO(2M) \times \prod_{i=1}^K SU(N_i)\times U(1)_i$   \\
\hline \hline
$(\Adj_{U(N_i)})_{0(i)}$ & $H^*(K3,V_i \otimes V_i^{\ast})$  \\
\hline
$(\Sym_{U(N_i)})_{2(i)}$ & $H^*(K3,\bigwedge^2 V_i)$  \\
$(\Anti_{U(N_i)})_{2(i)}$ & $H^*(K3, \bigotimes^2_s  V_i)$  \\
\hline
$(\N_i,\N_j)_{1(i),1(j)}$ & $H^*(K3, V_i \otimes V_j)$ \\
$(\N_i,\ov \N_j)_{1(i),-1(j)} $ &  $H^*(K3, V_i \otimes V_j^{\ast})$ \\
\hline
$(\Adj_{SO(2M)})_0$ & $H^*(K3,{\cal O})$ \\
$({\bf 2M}, \N_i)_{1(i)}$ & $H^*(K3, V_i)$\\
\hline
\end{tabular}
\caption{\small Perturbative massless spectrum with the structure group  taken to be 
$G=\prod_{i=1}^{K} U(n_i)$. 
The net number of hyper multiplets in complex representations 
is given by $-\chi(W)$ associated to the cohomology class $H^*(K3,W)$ as defined in~(\ref{RRH}). 
The massless spectrum contains also the supergravity sector, 20
neutral hypers encoding the $K3$ geometry and the universal tensor
multiplet.}
\label{TchiralSO32}
\end{center}
\end{table}
from which the anomaly polynomial in the perturbative smooth case is computed~\cite{Honecker:2006dt}, 
\bea
I^{SO(32)}_8 &=& \left(\tr R^2 +2 \, \tr_{SO(2M)} F^2 + 4 \sum_i (\ch_2(V_i) +n_i) \tr_{U(N_i)} F^2
\right)\times 
\nonumber\\
&& \times  \left(\tr R^2 - \tr_{SO(2M)} F^2 -2 \, \sum_i \, n_i \, \tr_{U(N_i)} F^2
\right)
\nonumber\\
&& 
+ \frac{1}{3} \left(\sum_i c_1(V_i) \tr_{U(N_i)} F\right)
\times  \left(\sum_j c_1(V_j) \left[
\tr_{U(N_j)} F \, \tr R^2 - 16  \tr_{U(N_j)} F^3 
\right]\right).
\nonumber
\eea
Comparing with the anomaly polynomials at the orbifold point~(\ref{Eq_Anomaly_Orbifold_SO32})
leads to the following observations:
\begin{itemize}
\item
In all orbifold cases where the $SO(2M_i)$ group occurs only with fundamental and not with spinor representations,
the coefficient $\alpha_i=2$ matches with the smooth case.
\item
The non-Abelian parts of the anomaly polynomial containing  $SU(N_i)$ gauge factors match when identifying the
orbifold and smooth parameters as follows,
\bea
\beta_i &=& 4 \, (\ch_2(V_i) +n_i), 
\nonumber\\
1 &=& n_i.
\nonumber
\eea
The second equation reveals that the background consists of {\it line bundles}  with structure group $U(1)$ embedded in $U(N_i)$, 
and from the first equation the second Chern characters of these line bundles can be determined using the orbifold data.
\item
The $SO(2M_i)$ gauge factors at the orbifold point with spinor representations in the
massless spectrum and $\alpha_i \neq 2$
are replaced by $U(M_i)$ factors on $K3$ with the identification of coefficients
\bea
SO(2M_i) &\rightarrow& U(M_i)
\nonumber\\
2 \, \alpha_i &=& 4 \, (\ch_2(V_i) +n_i), 
\nonumber\\
1 &=& n_i.
\nonumber
\eea
As for the previous case, in the smooth compactification a line bundle is embedded in $U(M_i)$ and the second Chern character 
of this line bundle can be determined from the corresponding coefficient $\alpha_i$ of the anomaly polynomial at 
the orbifold point.
\item
The $U(1)$ charges at the orbifold point cannot be reproduced by the smooth ansatz as one can easily see from the fact 
that, e.g.,  in the smooth case all fundamental representations of $SU(N_i)$ carry charge $1$ under $U(1)_i$ 
whereas in the orbifold limit their charge is $1+\frac{n}{m}$ in the $m^{th}$ twisted sector for some integer $n$ (or half-integer for spinorial shifts)  
and 1 in the untwisted sector. Similarly, antisymmetric representations in the smooth case and untwisted orbifold sector have $U(1)$ charge 2, in 
the $m^{th}$ twisted orbifold sector the charges are $2+\frac{n'}{m}$.\\
Furthermore, at the orbifold point, the complete anomaly polynomial factorizes as $4\times 4$ which suggests that the couplings 
to the (sometimes not even as twisted singlets identifiable) $c^{(4)}_k$ are absent. 
The Green-Schwarz counter term involves only the supergravity and the universal tensor 
multiplet,\footnote{The anti-self dual combination of $b^{(2)}_0$ and its dual $c^{(2)}_0$ forms part of the supergravity multiplet, whereas
the self dual combination is the tensor of the universal tensor multiplet.}
 and thus the $U(1)$ at the orbifold point is expected to remain massless.\footnote{In principle the $2 \times 6$ term could also vanish due to 
zero couplings of the $b^{(0)}_k$ in the presence of mass terms. Whereas in the smooth case, the field theoretic
couplings are well understood~\cite{Honecker:2006dt}, a similar analysis for
the orbifold cases has not been performed in detail, and in cases such as the
$T^4/\Z_2$ models, there are non-Abelian singlets only in the untwisted sector,
all other potential candidates for $b^{(0)}_k$ fields are only identified 
in the blown-up phase.
} 
\end{itemize}
The above comparison shows that the {\it non-Abelian} charges of the orbifold compactifications can be reproduced by 
embedding line bundles $L_i$ in $U(N_i)$ gauge factors, and the general massless spectrum simplifies due to the absence of 
symmetric representations as shown in table~\ref{Tab:SO32_Match_Ls}. 
This statement is true up to the caveat that in the smooth case, sometimes
only smaller subgroups occur, e.g. $SU(M)$ gauge factors instead of $SO(2M)$ with spinor representations or $SU(N-1)$ 
instead of $SU(N)$ with third rank antisymmetric representations.
\begin{table}[htb!]
\renewcommand{\arraystretch}{1.2}
\begin{center}
\begin{tabular}{|c||c|c|}
\hline
\hline
reps. & \# Hyper & \# Vector
\\\hline
$(\Adj_{U(N_i)})_{0}$ & 0 & 1
\\
$(\Adj_{SO(2M)})_0$ & 0 & 1 
\\\hline
$(\Anti_{U(N_i)})_{2(i)}$ & $-\chi(L_i^2)$ & 0 
\\
$(\N_i,\N_j)_{1(i),1(j)}$ & $-\chi(L_i \otimes L_j)$ & 0 
\\
$(\N_i,\ov{\N}_j)_{1(i),-1(j)}$ & $-\chi(L_i \otimes L_j^{-1})$ & 0
\\
$({\bf 2M}, \N_i)_{1(i)}$ & $-\chi(L_i)$ & 0
\\\hline
\end{tabular}
\end{center}
\caption{Matching perturbative orbifold spectra by embedding line bundles $L_i$ in $\prod_i U(N_i)$.}
\label{Tab:SO32_Match_Ls}
\end{table}
The ``tadpole cancellation condition''~(\ref{tcc}) for this kind of embeddings is determined via
\bea
\tr \ov{F}^2 &=& \frac{1}{30} \Tr \ov{F}^2 = 2 \sum_i N_i \ov {F}^2_{U(1)_i}
= 16 \pi^2 \sum_i N_i \, \ch_2(L_i),
\nonumber\\
\tr \ov{R}^2 &=& 2 \, \tr^{SU(2)}_f \ov{R}^2 = -16 \pi^2 c_2(K3),
\nonumber
\eea
with $c_2(K3)=24$ and reads
\bea
\sum_i N_i \; \ch_2(L_i) = -24.
\label{tcc_SO32}
\eea
Equation~(\ref{tcc_SO32})
serves as a good check for the consistency of a model
constructed using the coefficients of the anomaly polynomial at the orbifold point to be explained below.

In analogy to the orbifold case, the embedding of a line bundle $L$ into a $U(N)$ subgroup of $SO(32)$ can be denoted
by
\bea
(\!\!\!&\underbrace{L,\ldots,L}&\!\!\!,0,\ldots,0).
\nonumber\\
&  N \times &  
\label{Eq:Id_Shift_SO}
\eea
Also in complete analogy to the orbifold case, the resulting spectrum is {\it independent of the sign} of the first Chern class of the line 
bundle,\footnote{This can be seen explicitly by computing the multiplicities of bifundamental states if $L$ is embedded in $U(N_1)$ and 
$L^{-1}$ in $U(N_2)$. In this case, $Q_{\rm massless} = N_2 Q_1 + N_1 Q_2$ is the charge of the massless Abelian gauge factor, and
$(\N_1,\N_2)_{1,1}$ has multiplicity $\chi({\cal O}) = 2$ which implies the existence of vector multiplets
in the bifundamental representation and thereby the gauge enhancement $SU(N_1) \times SU(N_2) \times U(1)_{\rm massless}\rightarrow SU(N_1+N_2)$.}
only the K-theory constraint changes. Therefore, all embeddings 
\bea
(\!\!\!&\underbrace{L,\ldots,L}&\!\!\!,\underbrace{L^{-1},\ldots,L^{-1}},0,\ldots,0).
\nonumber\\
& N_1 \times & \quad N_2 \times  
\nonumber
\eea
are equivalent, and their total first Chern class is
\bea
c_1(W_{total})= (N_1-N_2) \, c_1(L).
\label{E:SO_Ktheory}
\eea
The generalization to embeddings with several line bundles or different powers $L^n$ is straightforward.

\begin{table}[p]
\renewcommand{\arraystretch}{1.2}
\begin{center}
\begin{tabular}{|c||c||c|c||c|c||c|c||c|c|}
\hline\hline
\# & $2M$ & $N_0$ & $\!\!\ch_2(L_0)\!\!$ 
& $N_1$ & $\!\!\ch_2(L_1)\!\!$ 
 & $N_2$ & $\!\!\ch_2(L_2)\!\!$ 
 & $N_3$ & $\!\!\ch_2(L_3)\!\!$ 
\\\hline\hline
2a & 28 & - & - & 2 & -12 & (2) & (-4) & - & -
\\\hline
3,4,6a & 28 & - & - & 2 & -12 & - & - & - & -
\\\hline
\hline
2b & 20 & 6 & -4 & - & -  & - & -  & - & -
\\\hline
2c & - & - & - & 16$^{\star}$ & -1 & - & - & - & -
\\\hline\hline
3b & 20 & - & - & 5$^{\star}$ & -3 & - & - & - & -
\\\hline
3c & 16 & - & - & 8 & -3 & - & -  & - & -
\\\hline
3d & - & 5 & $-\frac{9}{4}$ & 11$^{\star}$ & $-\frac{3}{4}$ & - & -& - & -
\\\hline
3e & - & - & - & 14 & $-\frac{3}{4}$ & 2 &  $-\frac{27}{4}$ & (2) & 
($-\frac{9}{4}$)
\\\hline\hline
4a' & 28 & - & - & 2 & -8 & - & -  & - & -
\\\hline
4b & 24 & - & - & 2 & -6 & (2) & (-2) & 2 & -6 
\\\hline
4c & 20 & 4 & -2 & 2 & -8 & - & - & - & -
\\\hline
4d & 16 & 6 & -2 & 2 & -6 & - & - & - & -
\\\hline
4e & - & 6 & $-\frac{3}{2}$ & 10$^{\star}$ & $-\frac{3}{2}$ & - & - & - & -
\\\hline
4e' & - & 6 & $-\frac{5}{2}$ & 10 & $-\frac{1}{2}$ & - & - & - & -
\\\hline
4f & - & 4 & $-\frac{5}{2}$ & 10 & $-\frac{1}{2}$ & (2) & 
($-\frac{5}{2}$) & 2 &  $-\frac{9}{2}$
\\\hline
4g & 18 & - & - & 6$^{\star}$ & -3 & - & - & -  & -  
\\\hline
4h & 14 & - & - & 6$^{\star}$ & -3 & 4 & -2 & - & -
\\\hline
4i & - & 5$_{i=1,2}$ & -1$_{i=1,2}$ & 6$^{\star}$ & -2 & - & - & - & -
\\\hline
4j &  - & - & - & 14 & $-\half$ & - & - & - & -
\\\hline
4k &  - & - & - & 15 & $-\frac{3}{4}$ &- & - & - & -
\\\hline
4l & - & - & - & 11 & $-\frac{3}{4}$ & 5 & $-\frac{7}{4}$ & - & -  
\\\hline
4m & - & - & - & 9 & $-\frac{3}{4}$ & 7 & $-\frac{7}{4}$ & - & -  
\\\hline
4n & - & - & - & 13 & $-\frac{3}{4}$ & 3 & $-\frac{15}{4}$ & - & - 
\\\hline\hline
6b & - & $\begin{array}{c}6\\ 5\end{array}$ &  $\begin{array}{c} -\frac{3}{2}\\ -\half\end{array}$
& 5 & $-\frac{3}{2}$ & - & - & - & -
\\\hline
\end{tabular}
\end{center}
\caption{Determining second Chern characters from the anomaly polynomial according to equation~(\ref{Eq:SO32_Coeffs_Instanton_numbers}). 
$SU(N_0)$ corresponds to an $SO(2N_0)$ gauge group at the orbifolds with $\alpha_i \neq 2$ and spinor
representations in the spectrum.  
In some $T^4/\Z_{2N}$ cases, one of the $SU(2)$
gauge factors is not reproduced by the smooth ansatz. These factors are displayed in parenthesis.
Furthermore a star 
denotes the fact that representations unavailable in the smooth case
occur at the  orbifold point, e.g. a ${\bf 20}$ of $SU(6)$. 
In the smooth ansatz, an $SU(N-1)$ gauge factor occurs for these cases instead of the $SU(N)$ factor listed.}
\label{Tab:SO32factorisation_instanton}
\end{table}

\subsubsection{Matching of $SO(32)$ heterotic orbifold and $K3$ spectra}
\label{Subsec:SO_Explicit_Matches}
In this section, we use the general form of the spectrum for compactifications with 
line bundles in table~\ref{Tab:SO32_Match_Ls} in order to find smooth matches of the models at the orbifold point.

To start with, we use the relation
\bea
\left.
\begin{array}{c}
\beta_i \\
2 \, \alpha_i
\end{array}
\right\}
= 4 \left(\ch_2(L_i) +1 \right)
\label{Eq:SO32_Coeffs_Instanton_numbers}
\eea
for the coefficients of $SU(N_i)$ gauge factors (or $SO(2N_i)$ with spinor representations at the orbifold point) 
in the anomaly polynomial to determine the second Chern characters 
of various line bundles, and in the second step we check if these satisfy the tadpole cancellation condition~(\ref{tcc_SO32}).
The coefficients $\alpha_i$, $\beta_i$ at the orbifold point are listed in tables~\ref{Tab:SO32factorisation_coeffs_normal}
to~\ref{Tab:SO32factorisation_coeffs_exc_3}. 
The second Chern characters computed using the relation~(\ref{Eq:SO32_Coeffs_Instanton_numbers})
are displayed in table~\ref{Tab:SO32factorisation_instanton}. Apart from these, further line bundles embedded in
$U(1)$ gauge factors can occur. Their second Chern characters cannot be determined simply from the anomaly polynomial, but
from the tadpole cancellation condition and the multiplicity of fundamental representations of the non-Abelian gauge factors in 
the model in question.

\begin{table}[t]
\renewcommand{\arraystretch}{1.2}
\begin{center}
\begin{tabular}{|c||c|c||c|c||c|c|}
\hline
\hline
\# & $N_1$ & $\ch_2(L_1)$  
& $N_2$ & $\ch_2(L_2)$ 
& $N_3$ & $\ch_2(L_3)$ 
\\\hline\hline
2-6a & 2 &  -12 & - & - & - & -
\\
2b & 6 & -4  & - & - & - & -
\\
2c & 15 & -1 & 1 & -9 & - & - 
\\\hline
3b & 4 & -3 & 1 & -12 & - & -
\\
3c & 8 & -3 & - & - & - & -
\\
3d & 10 &  $-\frac{3}{4}$ & 5 & $-\frac{9}{4}$ & 1 & $-\frac{21}{4}$  
\\
3e & 14 & $-\frac{3}{4}$ & 2 & $-\frac{27}{4}$ & - & -
\\\hline
4b & 2 & -6 & 2 & -6 & - & -
\\
4c & 2 & -8 & 4 & -2 & - & -
\\
4d & 2 & -6 & 6 & -2 & - & - 
\\
4e & 6 & $-\frac{3}{2}$ & 10 & $-\frac{3}{2}$ & - & -
\\
4e' & 6 & $-\frac{5}{2}$ & 9 & $-\half$ & 1 & $-\frac{9}{2}$ 
\\
4f & 4 & $-\frac{5}{2}$ & 10 & $-\half$ & 2 & $-\frac{9}{2}$ 
\\
4k & 15 & $-\frac{3}{4}$ & 1 & $-\frac{51}{4}$ & - & - 
\\\hline
\end{tabular}
\end{center}
\caption{Matching some perturbative $SO(32)$ orbifold spectra on $T^4/\Z_N$ for $N=2,3,4$. Due to the relation $\ch_2(L) = \half c_1(L)^2$ for a line bundle,
the second Chern characters are in general expected to be multiples of one-half. In the cases of spinorial shifts, an additional factor of 
one-half appears in the definition 
of the line bundle upon our identification~(\ref{Eq:Id_Shift_SO}) leading to multiples of $1/8$ for the second Chern characters. 
The shortest possible shift vectors for 3d, 3e are not the ones displayed in table~\ref{Tab:SO32_Pert_Orbifold_Spectra}, 
but are spinorial ones which are obtained by subtracting the spinorial weight $(\half, \ldots,\half)$ of $SO(32)$,
namely $-\frac{1}{6}(1^{10},-1,3^5)$ and $-\frac{1}{6}(1^{14},3^2)$, respectively.}
\label{Tab:SO32_Match_Chern_23}
\end{table}

\begin{table}[p]
\renewcommand{\arraystretch}{1.2}
\mbox{}\hspace{-.7cm}
\begin{tabular}{|c||c|c|}
\hline
\hline
\# & Gauge group & Matter
\\\hline\hline
2-6a & $SO(28) \times U(2)$ & $46 \, (\1,\1)_2 + 10 \, ({\bf 28},\2)_1 $ 
\\\hline
2b &  $SO(20) \times U(6)$ & $14 \, (\1,{\bf 15})_2 + 2({\bf 20},\6)_1$
\\\hline
2c & $U(15) \times U(1)$ & $2({\bf 105})_{2,0} +14({\bf 15})_{1,-1}+ 2({\bf 15})_{1,1}$
\\\hline\hline
3b  &  $SO(22) \times U(4) \times U(1)$ & $({\bf 22},\4)_{1,0} + 10 ({\bf
  22},\1)_{0,1} + 10(\1,\6)_{2,0} +25(\1,\4)_{1,1} + (\1,\4)_{1,-1}$  
\\\hline
3c &  $SO(16) \times U(8)$ & $10(\1,{\bf 28})_2 +  ({\bf 16},\8)_1$
\\\hline
3d &  $U(10) \times U(5) \times U(1)$ & 
$({\bf 45},\1)_{2,0,0} + 7 (\1,{\bf 10})_{0,2,0} $
\\
 & &
$+ 2 ({\bf 10},\5/\ov{\5})_{1,\pm 1} 
+ 8({\bf 10},\1)_{1,0,\pm 1} + 11 (\1,\5)_{0,1,\pm 1}$
\\\hline
3e & $U(14) \times U(2)$ & $({\bf 91})_{2,0} + 10({\bf 14},\2)_{1,1}+({\bf 14},\ov{\2})_{1,-1} + 25(\1,\1)_{0,2}$
\\\hline\hline
4b & $SO(24) \times U(2)^2$ & 
$4 ({\bf 24},\2,\1)_{1,0} + 4({\bf 24},\1,\2)_{0,1} $
\\
& & $+ 22 (\1,\1,\1)_{2,0} +  22 (\1,\1,\1)_{0,2}$
\\
& & $+20 (\1,\2,\2/\ov{\2})_{1,\pm 1}$
\\\hline
4c & $SO(20) \times U(2) \times U(4)$ &
$6 ({\bf 20},\2,\1)_{1,0} +16(\1,\2,\4/\ov{\4})_{1,\pm 1}$
\\
& & $6 (\1,\1,\6)_{0,2} + 30 (\1,\1,\1)_{2,0}$ 
\\\hline
4d & $SO(16) \times U(2) \times U(6)$ & 
$6 (\1,\1,{\bf 15})_{0,2} + (\1,\1,\1)_{2,0}$
\\
& & $12 (\1,\2,\6/\ov{\6})_{1,\pm 1} + 4 ({\bf 16},\2,\1)_{1,0}$
\\\hline
4e & $U(6) \times U(10)$ & 
$4 ({\bf 15},\1)_{2,0} + 4 (\1,{\bf 45})_{0,2} + 2({\bf 10},\6/\ov{\6})_{1,\pm 1}$
\\\hline
4e' & $U(6) \times U(9) \times U(1)$ &
$8 ({\bf 15},\1)_{2,0,0} + 2(\6,\9/\ov{\9},\1)_{1,\pm 1,0}$ 
\\
&& $+ 10 (\6,\1,\1)_{1,0,\pm 1} + 6 (\1,\9,\1)_{0,1, \pm 1}$
\\\hline
4f & $U(4) \times U(10) \times U(2)$ & 
$8 (\6,\1,\1)_{2,0,0} + 16 (\1,\1,\1)_{0,0,2}$+
\\
&& $2(\4,{\bf 10}/\ov{\bf 10},\1)_{1, \pm 1,0} + 10 (\4,\1,\2/\ov{\2})_{1,0,\pm 1}
+ 6 (\1,{\bf 10},\2/\ov{\2})_{0,1,\pm 1}$
\\\hline
4k & $U(15) \times U(1)$ & $({\bf 105})_{2,0} + 23 \, ({\bf 15})_{1, \pm 1}$ 
\\\hline
\end{tabular}
\caption{Massless spectra for smooth $SO(32)$ compactifications. The supergravity, universal
  tensor and twenty neutral hyper multiplets are not listed. Only the overall
  number of non-Abelian charges is counted, i.e. $16({\bf 15})_{1,\pm 1}$
  denotes a net number of 16 hyper multiplets transforming as ${\bf 15}$ with an arbitrary decomposition
  into $U(1)$ charge assignments $(1,1)$ or $(1,-1)$, $x ({\bf 15})_{1,1} + (16-x) ({\bf 15})_{1,-1}$. 
In the same spirit $\5/\ov{\5}$ means that the multiplet transforms
either in the fundamental or its conjugate representation and this difference is not specified by the second Chern characters, but the first Chern 
classes (which are not computable via the anomaly polynomial) of the line bundles are required.}
\label{Tab:SO32_Smooth_23}
\end{table}

Table~\ref{Tab:SO32factorisation_instanton} can be compared to the second Chern characters of some consistent
 smooth models listed in table~\ref{Tab:SO32_Match_Chern_23}
with their spectra displayed in table~\ref{Tab:SO32_Smooth_23}. 
The models fall into several categories:
\begin{itemize}
\item 
One clearly sees the matching of all non-Abelian charges for models 3-6a, 3c and 4k 
with the orbifold case. In the  cases 3-6a and 3c, just one line bundle is sufficient, whereas the matching of 4k requires two 
line bundles with different second Chern characters.
\item
For 2a, 3e and 4b the matching works nicely when an $SU(2)$ factor  at the orbifold point is ignored.
In the first two cases, one line bundle is embedded. For example  for 3e, $\ch_2(L_2) = 9 \, \ch_2(L_1)$ leads to the natural 
identification $L_2 = L_1^3$, which is the expected value using the shift vector of minimal length,
$\frac{1}{3}(1^{14},0^2) - (\frac{1}{2},\ldots,\frac{1}{2}) = -\frac{1}{6}(1^{14},3^2)$.\\
In case of model 4b, two independent line bundles with the same second Chern character are needed; otherwise $U(2)^2$ would be enhanced to $U(4)$. 
\item
Models 2b, 4c, 4d, 4e, 4f  match provided a breaking 
\bea
SO(2M) \rightarrow U(M)
\label{Eq:breakSO}
\eea
occurs (for the decompositions of representations see appendix~\ref{App:Decomp}), where 2b and 4c are realized by the embedding of 
one line bundle, and 4d, 4e, 4f require (at least) two different bundles.
In contrast to the other cases, the $SO(12)$ ($SO(8)$) gauge group in 4e (4f) stems from zero entries in the shift vector and the requirement of
an $U(6)$ ($U(4)$) factor instead can only be seen by the existence of the ${\bf 32}_+$ ($\8_{\pm}$) representation in the $T^2$ 
($T$ and $T^2$) sector at the orbifold point.
\item
2c and 3b match with one line bundle embedded 
upon the breaking
\bea
SU(N+1) \rightarrow SU(N) \times U(1)
\label{Eq:breakSU}
\eea
suggested by the number of identical entries in the shift vectors. For example, for 2c the identifications implies 
$\frac{1}{4}(1^{15},-3) \rightarrow (L, \ldots , L,L^{-3})$
(decompositions of representations are again given in appendix~\ref{App:Decomp}),
and the spectrum contains two hyper multiplets in the antisymmetric and 14+2=16 in the fundamental representation of $SU(15)$, while at the orbifold 
two antisymmetric and 16 fundamentals of $SU(16)$ appear.
\item
In cases 3d, 4e', both types of breakings~(\ref{Eq:breakSO}) and ~(\ref{Eq:breakSU}) are needed to find a smooth match. The breaking of the $SO(12)$
gauge group  in 4e' is again not visible from the zero entries of the shift vectors, and at least two different line bundles have to be embedded in each model.
\item
For model 4a', smooth matches are possible with two line bundles embedded in $SO(28) \times U(1)^2$, but since 
$\sum_{i=1}^2 \ch_2(L_i) = -24$  is the only model building constraint, we do not display any spectrum.
Similarly, model 4j should have a smooth match with $L_1,L_2,L_3$ embedded in $U(14) \times U(1)^2$.
\item
Models 4g, 4h, 4i require some $SU(6) \rightarrow SU(5) \times U(1)$ breaking due to the existence of ${\bf 20}$ representations in the $T^2$ sector
which is not evident in the shift vector with six identical entries, similarly 
smooth matches to 4l, 4m, 4n require some breaking of the type~(\ref{Eq:breakSU}) due to the existence of fields transforming only 
in the fundamental, not bifundamental representation of the non-Abelian gauge factors.
6b can only be matched provided both orthogonal groups are broken along the lines of equation~(\ref{Eq:breakSO}) and afterwards some breaking 
of the type~(\ref{Eq:breakSU}) occurs.
\end{itemize}

In summary, for ten models we find smooth matches with just one line bundle embedded in $SO(32)$ fitting with the identification
of shift vectors and line bundle embeddings. In the other cases, the smooth matches are more involved. For example, the other explicitly 
worked out $T^4/\Z_4$ matches require two independent line bundles, which suggests a correspondence between the number of orbifold 
twist sectors and line bundles embedded.


\subsection{$U(1)$ bundles inside $E_8$}
\label{Subsec:E8_K3_U1s}

In section~\ref{Subsec:SO32bundles}, we have shown that the shift vector of the $SO(32)$ heterotic orbifold has a direct interpretation 
in terms of the embedding of  line bundles. In order to transfer the argument to the $E_8 \times E_8$ case, we consider the following 
successive breaking
\bea
\begin{array}{c}
E_8 \\
{\bf 248}
\end{array}
&\rightarrow&
\begin{array}{c} 
SO(16) \\
\left(
\begin{array}{c}
{\bf 120} \\
{\bf 128}
\end{array}
\right)
\end{array}
\rightarrow 
\begin{array}{c}
SO(14) \times U(1)\\
\left(
\begin{array}{c}
 ({\bf 91})_0 + (\1)_0 + [({\bf 14})_2 + c.c.] \\
 ({\bf 64})_1 + c.c.
\end{array}
\right)
\end{array}
\nonumber\\
\nonumber\\
\nonumber\\
& \rightarrow &
\begin{array}{c} 
SO(12) \times U(1)^2\\
\left(
\begin{array}{c}
({\bf 66})_{0,0} + 2(\1)_{0,0}\\
+[({\bf 12})_{2,0} + ({\bf 12})_{0,2} + (\1)_{2,2} + (\1)_{2,-2} + c.c.]\\
({\bf 32}_+)_{1,1} + ({\bf 32}_-)_{1,-1} +c.c.
\end{array}
\right)
\end{array}
\rightarrow \ldots.
\nonumber
\eea
These breakings can be cast into the compact notation $(2 \leqslant N \leqslant 6)$
\bea
E_8 &\rightarrow& SO(2N) \times U(1)^{8-N}
\nonumber\\
{\bf 248}  &\rightarrow& 
({\bf Adj}_{SO(2N)})_0 + (8-N) \times (\1)_0 
+ ({\bf 2N})_{\underline{\pm 2,0^{7-N}}}
+ (\1)_{\underline{\pm 2, \pm 2, 0^{6-N}}} 
\nonumber\\
&&
+\sum_{k=0}^{[4-\frac{N}{2}]} (\2^{N-1}_+)_{\underline{1^{8-N-2k},-1^{2k}}}
+\sum_{k=0}^{[\frac{7-N}{2}]} (\2^{N-1}_-)_{\underline{1^{7-N-2k},-1^{1+2k}}} , 
\nonumber\\
\label{Eq:E8_decomp}\\
\nonumber\\
E_8 &\rightarrow& U(1)^8
\nonumber\\
{\bf 248}  &\rightarrow& 8 (\1)_0 + (\1)_{\underline{\pm 2, \pm 2, 0^6}} 
+ \sum_{k=0}^4 (\1)_{\underline{1^{8-2k},-1^{2k}}} , 
\nonumber
\eea
where underlining of the charges denotes all possible permutations.
The $U(1)$ charge assignments serve, analogously to  the $SO(32)$ case,  as a guideline to the correct assignment of bundles,
namely the $i^{th}$ charge entry $n_i$ in~(\ref{Eq:E8_decomp}) corresponds to the bundle $L_i^{n_i}$.

The tadpole contribution for all successive breakings is computed from
\bea
\tr_{E_8} \ov{F}^2 = \frac{1}{30} {\rm Tr}_{E_8} \ov{F}^2 =16 \pi^2 \cdot 4 \sum_{i=1}^N \ch_2(L_i) 
\nonumber
\eea
with all charge assignments integer valued and the corresponding integer powers of line bundles associated as for the following example.

Consider for concreteness the embedding of two line bundles $L_1$, $L_2$ inside an $E_8$ factor,
\bea
(L_1, L_2, 0^6).
\eea
The resulting massless spectrum consists of the gauge group $SO(12) \times U(1)^2_{\rm massive}$ and matter with the
multiplicities listed in table~\ref{E8:Lassignments}.
\begin{table}[htb]
\renewcommand{\arraystretch}{1.2}
\begin{center}
\begin{tabular}{|c|c|}
\hline
\hline
\# & rep.
\\\hline\hline
$-\chi(L_1^2)$ & $({\bf 12})_{2,0}$
\\
$-\chi(L_2^2)$ & $({\bf 12})_{0,2}$
\\
$-\chi(L_1 \otimes L_2)$ & $({\bf 32}_+)_{1,1}$
\\
$-\chi(L_1 \otimes L_2^{-1})$ & $({\bf 32}_-)_{1,-1}$
\\
$-\chi(L_1^2 \otimes L_2^2)$ & $(\1)_{2,2}$
\\
$-\chi(L_1^2 \otimes L_2^{-2})$ & $(\1)_{2,-2}$
\\\hline
\end{tabular}
\end{center}
\caption{Matter multiplicities for $L_1$, $L_2$ embedded in $SO(12) \times U(1)^2 \subset E_8$.}
\label{E8:Lassignments}
\end{table}

Similar to the $SO(32)$ embeddings of line bundles, non-Abelian gauge enhancements occur for special combinations 
of several line bundles. Consider for concreteness the case which corresponds to
the so called orbifold ``standard embedding''. 
The spectrum in table~\ref{E8:Lassignments} contains states $({\bf 32}_{\mp})_{1,\mp 1}$ and $(\1)_{2,\mp 2}$ 
transforming as the trivial line bundle ${\cal O}$ for $L_2 = L_1^{\pm 1}$ thereby leading to the
gauge enhancement
\bea
SO(12) \times U(1)_{\rm massless} \times U(1)_{\rm massive}  \rightarrow SO(12) \times SU(2) \times U(1)_{\rm massive} \rightarrow E_7 \times U(1)_{\rm massive},
\eea
where in the first step, one observes that $U(1)_{\rm massless}=U(1)_1 \mp U(1)_2$ remains massless 
due to the linear dependence of the two line bundles,
and the states in the $(\1)_{2,\mp 2}$ representation lead to  $U(1)_{\rm massless} \rightarrow SU(2)$. 
In the second step, the vectors transforming as $({\bf 32}_{\mp},\2)_0$ provide the $E_7$ enhancement.

The two choices $L_2=L_1^{\pm 1}$ differ only in their fulfillment of the K-theory constraint,
\bea
c_1(W_{total}) = c_1(L_1) + c_1(L_2) = \left\{\begin{array}{cc}
2 \, c_1(L_1) & L_2 = L_1 \\
0 & L_2 = L_1^{-1}
\end{array}\right.
\nonumber
\eea
in analogy to the $SO(32)$ case~(\ref{E:SO_Ktheory}).

The resulting multiplicities of the spectrum are listed in table~\ref{Tab:E_8_SS_1}. 

The same line of argument applies to other cases, those who are relevant for the matching of $T^4/\Z_N$ orbifold spectra 
are  listed  in tables~\ref{Tab:E_8_SS_1},~\ref{Tab:E_8_SS_2} and~\ref{Tab:E_8_SS_3}. 
Notice that these tables are valid also for Calabi-Yau compactification, since
the dimension of the space affects only the expansion of $\chi(L)$, which for four dimensional models is 
$\chi(L)_{CY_3} = \int_{CY_3} \left(\ch_3(L) + \frac{1}{12} c_2(CY_3) \, c_1(L) \right)$.

\begin{table}[htb]
\renewcommand{\arraystretch}{1.2}
{\footnotesize
\mbox{}\hspace{-.5cm}
\begin{tabular}{|c||c||c|c|c|c|}
\hline
\hline
 &  & $SO(14) \times U(1)$ & $E_7 \times U(1)$ & $E_6 \times SU(2) \times U(1)$ & 
$SU(8) \times U(1)$
\\\hline\hline
$\!\!$\#V$\!\!$& \#H 
&  $\!\!\!\begin{array}{c}(L,0^7) \\ (L^{1/4},\ldots, L^{1/4},L^{-3/4})\end{array}\!\!\!$  
&$\!\!\!\begin{array}{c}(L^{1/2}, L^{1/2},0^6)\\(L^{1/4},\ldots, L^{1/4}) \end{array}\!\!\!$
&  $\!\!\!\begin{array}{c}(L^{1/2}, L^{1/2},L,0^5) \\(L^{1/2}, \ldots,L^{1/2},0^2)  \end{array}\!\!\!$ 
&  $\!\!\!\!\begin{array}{c} (L^{1/2}, L^{1/2}, L^{1/2},L^{1/2}, L, 0^3)\!\!\!\!\\
(L^{1/2}, \ldots,L^{1/2},L^{-1/2}) \\
 (L^{1/4},\ldots, L^{1/4},L^{5/4})\end{array} \!\!$  
\\\hline\hline
1 & 0 & $({\bf 91})_0$ & $({\bf 133})_0$ & $({\bf 78},\1)_0$ & 
$({\bf 63})_0$
\\\hline
1 & 0 & - & - & $(\1,\3)_0$ & - 
\\\hline
1 & 0 & $(\1)_0$ & $(\1)_0$ &  $(\1)_0$ & $(\1)_0$
\\\hline
0 &$\!\!-\chi(L)\!\!$& $({\bf 64})_1$ & $({\bf 56)}_1$ & $(\ov{\bf 27},\2)_1$ &  
$({\bf 56})_1$
\\\hline
0 &$\!\!-\chi(L^2)\!\!$& $({\bf 14})_2$ & $(\1)_2$ & $({\bf 27},\1)_2$ & 
$({\bf 28})_2$
\\\hline
0 &$\!\!-\chi(L^3)\!\!$&  - & - & $(\1,\2)_3$ & 
$(\8)_3$
\\\hline
\end{tabular}
\caption{General spectrum from embedding a line bundle $L$ inside one $E_8$ factor. The embedding is specified by the shift vector
corresponding to the Cartan generators. In all cases, where the shift vector contains at least one zero entry, the signs of the 
exponents of the line bundle entries are arbitrary, and for shortness only positive exponents are listed. The powers of the line bundles
are rescaled such that the smallest multiplicity of a matter state is computed from $-\chi(L)$.
Part 1.}
\label{Tab:E_8_SS_1}}
\end{table}

%

\begin{table}[htb]
\renewcommand{\arraystretch}{1.2}
{\footnotesize
\mbox{}\hspace{-.7cm}
\begin{tabular}{|c||c||c|c|c|c|c|}
\hline
\hline
 &  &$\!\!\!SO(10) \times SU(3) \times U(1)\!\!\!$&$\!\!\!\!\!\!\!SU(7) \times SU(2) \times U(1)\!\!\!\!\!\!\!$&$\!\!\!SU(5) \times SU(3) \times SU(2) \times U(1)\!\!\!$
&$\!\!\!SO(12) \times U(1)^2\!\!\!$
\\\hline\hline
$\!\!\!$\#V$\!\!\!$& \#H & $(L,L,L,0^5)$ 
&$\!\!\!\!\!\!\!\begin{array}{c} (L^{1/2},\ldots,L^{1/2},L^{3/2},0^2) \\ (L^{1/4},\ldots,L^{1/4},L^{-7/4})\end{array}\!\!\!\!\!\!\!$
&$(L^{1/2},\ldots,L^{1/2},L^{5/2},0^2)$
&$\!\!\!(L^{1/2},L^{3/2},0^6)\!\!\!$
\\\hline\hline
1 & 0 & $({\bf 45},\1)_0$ & $({\bf 48},\1)_0$ & $({\bf 24},\1,\1)_0$ 
&$({\bf 66})_{0,0}$
\\\hline
1 & 0 & $(\1,\8)_0$ & $(\1,\3)_0$ & 
$\begin{array}{c} (\1,\8,\1)_0 \\(\1,\1,\3)_0 \end{array}$
&$(\1)_{0,0}$
\\\hline
1 & 0 & $(\1)_0$ &  $(\1)_0$ &  $(\1)_0$ 
&$(\1)_{0,0}$
\\\hline
0 & $\!\!\!\!-\chi(L)\!\!\!\!$ & $(\ov{\bf 16},\ov{\3})_{1}$ & $({\bf 21},\2)_1$ & $(\5,\3,\2)_1$ 
&$\begin{array}{c} ({\bf 12})_{3,1} \\ ({\bf 32}_-)_{-2,1} \end{array}$
\\\hline
0 & $\!\!\!\!-\chi(L^2)\!\!\!\!$ & $({\bf 10},\3)_2$ & $(\ov{\bf 35},\1)_2$ & $({\bf 10},\ov{\3},\1)_2$
&$\begin{array}{c}(\1)_{-4,2} \\ ({\bf 32}_+)_{1,2}\end{array}$
\\\hline
0 & $\!\!\!\!-\chi(L^3)\!\!\!\!$ & $({\bf 16},\1)_3$ & $(\ov{\7},\2)_3$ & $(\ov{\bf 10},\1,\2)_3$
& $({\bf 12})_{-1,3}$
\\\hline
0 & $\!\!\!\!-\chi(L^4)\!\!\!\!$ &  $(\1,\ov{\3})_4$ & $(\7,\1)_4$ & $(\ov{\5},\3,\1)_4$
& $(\1)_{2,4}$
\\\hline
0 & $\!\!\!\!-\chi(L^5)\!\!\!\!$ & - & - & $(\1,\ov{\3},\2)_5$
& - 
\\\hline
0 & $\!\!\!\!-\chi(L^6)\!\!\!\!$ & - & - & $(\5,\1,\1)_6$
& -
\\\hline
\end{tabular}
}
\caption{Embedding a line bundle $L$ inside  $E_8$. Part 2.
In the last column, the charge assignments are $(Q_{\rm massless},Q_{\rm massive})=(\frac{3Q_1-Q_2}{2},\frac{Q_1+3Q_2}{2})$
in terms of the original charges.}
\label{Tab:E_8_SS_2}
\end{table}

\begin{table}[htb]
\renewcommand{\arraystretch}{1.2}
{\footnotesize
\begin{center}
\begin{tabular}{|c||c||c|||c||c|}
\hline
\hline
\#V & \#H & $SU(5) \times SU(4) \times U(1)$ & \#H & $\begin{array}{c} (L,L,L,L,L,0^3) \\ (L^{1/2},L^{1/2},L^{1/2},L^{1/2},L^2,0^3)\end{array}$
\\\hline\hline
1 & 0 & $({\bf 24},\1)_0$ &  $-\chi(L^3)$ &  $(\5,\ov{\4})_3$
\\\hline
1 & 0 & $(\1,{\bf 15})_0$ &  $-\chi(L^4)$ & $({\bf 10},\1)_4$
\\\hline
0 & $-\chi(L)$ & $({\bf 10},\4)_1$ &   $-\chi(L^5)$ & $(\1,\4)_5$
\\\hline
0 & $-\chi(L^2)$ & $(\5,\6)_2$ & & 
\\\hline
\end{tabular}
\end{center}
}
\caption{Embedding a line bundle $L$ inside $E_8$. Part 3.}
\label{Tab:E_8_SS_3}
\end{table}

With the help of the counting of multiplicities in tables~\ref{Tab:E_8_SS_1} to~\ref{Tab:E_8_SS_3}, 
the general form of the non-Abelian part of the anomaly polynomial can be computed to be of the shape~(\ref{E8:Anomaly_Polynomial_General}) 
with the coefficients $a_r$ listed in table~\ref{Eq:E8_smooth_coeffs} and $\tilde{a}_r$ as in the orbifold case~(\ref{Eq:coeff_E8_atilde}).
\begin{table}[htb]
\renewcommand{\arraystretch}{1.2}
\begin{center}
\begin{tabular}{|c|c|}
\hline
\hline
$G_r$ & $a_r $
\\\hline\hline
$SO(12)$ & $-(5 \ch_2(L) + 6)$
\nonumber\\
$SO(14)$ & $-2(\ch_2(L)+3)$
\nonumber\\
$E_7$ & $-(\frac{1}{6} \ch_2(L)+1)$
\nonumber\\
$(E_6,SU(2))$ & $-(\ch_2(L)+2) \times (1,6)$
\nonumber\\
$SU(8)$ & $-4 (2\ch_2(L)+3)$ 
\nonumber\\
$(SO(10),SU(3))$ & $- 6\, (\ch_2(L)+1) \times (1,2)$
\\
$(SU(7),SU(2))$ & $-(14 \ch_2(L) + 12 ) \times (1,1)$
\\
$(SU(5), SU(3) ,SU(2))$ & $-6 \, (5 \ch_2(L) + 2) \times (1,1,1)$
\\
$(SU(5),SU(4))$ & $- (20 \ch_2(L) + 12) \times (1,1)$
\\\hline
\end{tabular}
\end{center}
\caption{Relation of coefficients in the $E_8 \times E_8$ anomaly polynomials and second Chern characters.}
\label{Eq:E8_smooth_coeffs}
\end{table}
As for the $SO(32)$ matchings, the coefficients $a_r$ of the non-Abelian gauge factors will serve in section~\ref{Subsec:E8matches}
as the guideline to compute the second Chern characters of the smooth models from the orbifold data, whereas the role of the $U(1)$ factors 
at the orbifold point cannot be recovered by the smooth ansatz.

\subsubsection{Explicit $K3$ realizations of $E_8 \times E_8$ orbifold spectra}
\label{Subsec:E8matches}

In this section, we give an explicit matching of some $E_8 \times E_8$ orbifold spectra by smooth $K3$ compactifications 
with one line bundle and comment on obstructions for other cases.

As for the $SO(32)$ case, the natural identification of $\Z_N$ orbifold shift vectors and line bundles is
\bea
\frac{1}{N} (1,\ldots,1,n,0,\ldots,0) \rightarrow
(L,\ldots, L, L^n, 0,\ldots,0),
\nonumber
\eea
and in order to compute the second Chern characters of the line bundles, the correspondence between orbifold
coefficients $a_r$ of non-Abelian gauge factors in the anomaly polynomial and second Chern characters in table~\ref{Eq:E8_smooth_coeffs}
is used. The result is obtained from the larges non-Abelian gauge factor and
listed in the fourth column of table~\ref{Tab:E8factorisation_systematics} in section~\ref{Sec:Ano},
where the powers of the line bundle $L$ 
from which the Chern characters are computed are those given in tables~\ref{Tab:E_8_SS_1} to~\ref{Tab:E_8_SS_3}.
The resulting values for $\ch_2(L)$ take the form of fractional numbers for model IIIe and several of those based on $T^4/\Z_N$ 
orbifolds for $N=4,6$. 
Those are the ones which have hyper multiplets in the twisted orbifold sectors with charges under the remainders of {\it both} $E_8$ gauge factors
and cannot be reproduced by our smooth ansatz with just one line bundle inside each $E_8$. 
The embeddings which have smooth matches along our general simple rules are given in table~\ref{Tab:E8_SS_Embeddings}
with their spectra in table~\ref{Tab:E8_SS_v2}.

As for the $SO(32)$ cases, there is a perfect match of all non-Abelian charges for models III-VIa and IIIb, and for IIa and IVb the massless spectra agree 
up to an additional $SU(2)$ gauge factor at the orbifold point. Models IIIc and IIId match at the non-Abelian level when decomposing 
$SU(N+1) \rightarrow SU(N) \times U(1)$, 
and  IVc matches upon the breaking $SO(16) \rightarrow SO(14) \times U(1)$.\footnote{The matching for this breaking is best seen by comparing 
the (for this model vanishing)
number of spinor representations: ${\bf 128} \rightarrow {\bf 64} + \ov{\bf 64}$. As displayed in appendix~\ref{App:Decomp}, the adjoint representation of 
higher rank contains two fundamental representation which together with two hyper multiplets group into massive vectors upon symmetry breaking. Therefore 8-2=6 is the 
correct number of {\bf 14}'s in the smooth model.} 
Model IIb requires the same breaking, and additionally the $SU(2)$ gauge factor is only 
present at the orbifold point.
Finally, IVe has a smooth match, but the $SU(2)$ representations don't agree. This mismatch is due to the mixing with a non-perturbative $SU(2)$ symmetry 
of the orbifold background~\cite{Kaplunovsky:1999ia}.

Except for IVe, all second Chern characters in table~\ref{Tab:E8_SS_Embeddings} are consistent with embedding (multiples of) the same line bundle 
in both $E_8$ gauge factors. Correspondingly, one linear combination of the Abelian gauge groups will stay massless. The $U(1)$ charges given in 
table~\ref{Tab:E8_SS_v2} are, however, the original ones. As an example, in model IVb, the massive and massless charges are proportional to $(Q_1 \pm Q_2)$.

\begin{table}[htb]
\renewcommand{\arraystretch}{1.2}
\begin{center}
\begin{tabular}{|c||c||c|c|}
\hline
\hline
\# & Embedding & $\ch_2(L)$ & $\ch_2(\tilde{L})$
\\\hline\hline
II-VIa & $(L^{1/2},L^{1/2},0^6; 0^8)$ & -12 & 0
\\\hline\hline
IIb & $(L,0^7;\tilde{L}^{1/2},\tilde{L}^{1/2},0^6)$ & -4 & -4  
\\\hline\hline
IIIb & $(L,0^7;\tilde{L},0^7)$ & -3 & -3
\\\hline
IIIc &  $(L^{1/2},L^{1/2},L^{1/2},L^{1/2},L,0^3;0^8)$ & -3 & 0
\\\hline
IIId &  $(L^{1/2},L^{1/2},L,0^5;\tilde{L}^{1/2},\tilde{L}^{1/2},0^6)$ & -3 & -3 
\\\hline\hline
IVb &  $(L^{1/2},L^{1/2},0^6;\tilde{L}^{1/2},\tilde{L}^{1/2},0^6)$ & -6 & -6 
\\\hline
IVc &  $(L^{1/2},L^{1/2},0^6;\tilde{L},0^7)$ & -8 & -2
\\\hline
IVe & $(L^{1/2},L^{1/2},L,0^5;\tilde{L},0^7)$ & -2 & -3
\\\hline
\end{tabular}
\end{center}
\caption{Line bundle embeddings for some smooth matches of orbifold spectra.
}
\label{Tab:E8_SS_Embeddings}
\end{table}

\begin{table}[htb]
\renewcommand{\arraystretch}{1.2}
\mbox{}\hspace{-.6cm}
\begin{tabular}{|c||c|c|}
\hline
\hline
\# 
& Gauge Group & Spectrum
\\\hline\hline
II-VIa & $E_7 \times U(1) \times E_8$ 
& $10 ({\bf 56})_{2} + 46(\1)_{4}$
\\\hline\hline
IIb & $SO(14) \times U(1) \times E_7  \times U(1)$ 
& $14({\bf 14},\1)_{2,0} +2({\bf 64},\1)_{1,0}$ 
\\
&  & $+2 (\1,{\bf 56})_{0,1} +14(\1)_{0,2}$
\\\hline\hline
IIIb  & $SO(14) \times  U(1) \times SO(14) \times  U(1)$ 
& $({\bf 64},\1)_{1,0} + (\1,{\bf 64})_{0,1}$
\\
& & $ +10 ({\bf 14},\1)_{2,0} +10  (\1,{\bf 14})_{0,2}$
\\\hline
IIIc & $ SU(8) \times U(1) \times E_8$ 
& $({\bf 56})_{1} + 10 ({\bf 28})_{2} + 25 (\8)_{3}$
\\\hline
IIId & $E_6 \times SU(2)  \times U(1) \times E_7 \times U(1) $
& $ ({\bf 27},\2;\1)_{1,0} + 10  ({\bf 27},\1;\1)_{2,0} + 25 (\1,\2;\1)_{3,0}$
\\
&&
$+(\1,\1;{\bf 56})_{0,1} + 10(\1)_{0,2}$
\\\hline\hline
IVb & $E_7 \times U(1) \times E_7 \times U(1)$ 
& $4 ({\bf 56};\1)_{1,0} + 4 (\1;{\bf 56})_{0,1} + 22 (\1)_{2,0} +22(\1)_{0,2}$
\\\hline
IVc &  $E_7 \times U(1) \times SO(14) \times U(1)$ 
& $6({\bf 56};\1)_{1,0} + 30 (\1)_{2,0} + 6 (\1;{\bf 14})_{0,1}$
\\\hline
IVe & $E_6 \times SU(2) \times U(1) \times  SO(14) \times U(1)$ 
& $6 ({\bf 27},\1;\1)_{2,0} + 16 (\1,\2;\1)_{3,0}$ 
\\
&& $+ (\1,\1;{\bf 64})_{0,1} + 10 (\1,\1;{\bf 14})_{0,2}$
\\\hline
\end{tabular}
\caption{Some perturbative smooth $E_8 \times E_8$ spectra. 
}
\label{Tab:E8_SS_v2}
\end{table}

Similarly to the $SO(32)$ matches, one line bundle is not sufficient for many $T^4/\Z_N$ models with $N=4,6$. 
As another example besides IVe, consider the shift vector $\frac{1}{4}(1,3,0^6)$. The ansatz $(L^{1/2}, L^{3/2},0^6)$ does not provide
the correct number of ${\bf 32}_+$ and ${\bf 32}_-$ spinor representations of $SO(12)$. Instead, by comparison with the 
multiplicities of the fundamental and spinor representations in the
orbifold spectra we obtain
the following constraints on two different line bundles embedded as in table~\ref{E8:Lassignments}:
\bea
\begin{array}{|c||c|c|c|c|c|}
\hline
& \mbox{IVg} & \mbox{IVh} & \mbox{IVi} & \mbox{IVl}  & \mbox{VIb} 
\\\hline\hline
 \ch_2(L_1) +\ch_2(L_2) & -6 & -5 & -4 & -\frac{5}{2} & -3 
\\\hline
c_1(L_1) \, c_1(L_2) & 0 & -3 & -2 & -\half & 0
\\\hline
\end{array}
\nonumber
\eea
Model IVg has then gauge group $SO(12) \times U(1)^2_{\rm massive} \times E_8$ with 
$4 ({\bf 32}_+)_{1,1} + 4 ({\bf 32}_-)_{1,-1} + x ({\bf 12})_{2,0} + (20 - x) ({\bf 12})_{0,2} + 22 (\1)_{2,2} + 22 (\1)_{2,-2}$,
where $x$ depends on  the value of $\ch_2(L_1)$. Models IVh, IVi and VIb nicely work along the same lines.
For the remaining cases, the second Chern characters in table~\ref{Tab:E8factorisation_systematics}
are fractional and the matching is more complicated involving also splittings into several line bundles for other shift vectors.

In general, the gauge symmetry breaking is more involved than in model IVg where one has just an additional massive $U(1)$ factor. For example, 
the assignment of two line bundles $(L_1,L_1,L_2,0^5)$ leads to the gauge group $SO(10) \times SU(2) \times U(1)^2_{\rm massive}$.
For $L_2 = L_1$, the second factor is enhanced to $SU(3)$ as displayed in table~\ref{Tab:E_8_SS_2}, whereas for $L_2 = L_1^2$ 
the first factor is enhanced to $E_6$ as shown in table~\ref{Tab:E_8_SS_1}.

Since $T^4/\Z_N$ orbifolds with $N=4,6$ have more than one twist sector and more than one kind of fixed point, it is not surprising
that more than one line bundle is required to obtain smooth matches.


\section{Towards explicit realizations of line bundles}
\label{Sec:Expl_Line}

Throughout this article, we are working with line bundles whose second Chern characters are determined via the matching of the 
anomaly polynomials and the tadpole cancellation constraint. In this section, we speculate on the explicit realization of these line bundles.
The naive starting point is motivated by S-duality with type II compactifications, namely in the orbifold limit $T^4/\Z_2$, 
one starts with 
\bea
c_1(L) = \half \sum_{i=1}^{16} E_i
\quad
\Rightarrow
\quad
\ch_2(L)= -4,
\label{Eq:Z2_bundle}
\eea
where $i$ labels the orbifold fixed points and $E_i$ with $E_i \cdot E_j = - 2 \delta_{ij}$ the two forms associated to the
 blown down two-cycles at the orbifold point. 
The background gauge field is localized at the orbifold fixed points
and  democratically distributed among them.
The bundle ansatz~(\ref{Eq:Z2_bundle}) gives the correct second Chern character for models 2b and IIb.
Since 2c is based on a spinorial shift, one can speculate that the factor one-half directly enters the definition of the line bundle, 
\bea
c_1(L_{\rm 2c}) = \frac{1}{4}  \sum_{i=1}^{16} E_i,
\nonumber
\eea
which indeed gives the desired second Chern character.

In the same spirit, one can make the ansatz for a democratic distribution over the nine $T^4/\Z_3$ fixed points,
\bea
c_1(L) = \frac{1}{3} \sum_{i=1}^9 \left(E^{(1)}_i - E^{(2)}_i  \right)
\quad
\Rightarrow
\quad
\ch_2(L) = -3,
\eea 
with the intersection form equal to minus the Cartan matrix of $A_2$, i.e.
$E^{(1)}_i \cdot E^{(1)}_j = E^{(2)}_i \cdot E^{(2)}_j = - 2 \delta_{ij}$, $E^{(1)}_i \cdot E^{(2)}_j = \delta_{ij}$.
This fits with the second Chern characters of 
 models 3b, 3c, IIIb - IIId, and again one finds the matching with the spinorial shift 3e upon multiplication by one-half.
The value for the standard embeddings 3a and IIIa is obtained by multiplying by two.

The cases $T^4/\Z_N$ for $N=4,6$ are more involved due to the occurrence of different types of exceptional cycles at various
singularities as discussed in section~\ref{Sec:Orbifolds}.

Even though this ansatz fits nicely for several models, it is not at all obvious that this is indeed the correct correspondence since 
the partition functions mix contributions from the space-time and gauge sector embeddings.
This it what makes the transfer of our $T^4/\Z_N$ ansatz to $T^6/\Z_N$ compactifications
so difficult: whereas $K3$ is unique, out of the multitude of 
Calabi-Yau threefolds it is not clear if the blown-up model will have the same Hodge numbers 
as the  $T^6/\Z_N$ orbifold background.


\section{Flat directions and blow-up of the orbifold models}
\label{Sec:Blowup}
In the previous sections, the matching between orbifold models
and smooth K3 compactifications has been shown.
The matching is satisfactory only up to some caveats,
summarized in section~\ref{Subsec:SO_Explicit_Matches} for the
$SO(32)$ case (the $E_8$ case can be treated in a similar
fashion as discussed in~\ref{Subsec:E8matches}). The main point is a reduction of the gauge symmetry
in the matching, of the type  $SU(N)\rightarrow SU(N-1)\times U(1)$
or $SO(2N)\rightarrow SU(N)\times U(1)$, accompanied by the
reduction of the total rank of the gauge group due to the ``disappearance''
of a number of $U(1)$ factors equal to the number of bundles
introduced on the smooth model building side of the matching.
From the latter perspective these $U(1)$ factors are anomalous
and get a mass term due to a modified Green-Schwarz mechanism.

Such a gauge symmetry reduction is completely natural.
Indeed, the matching between orbifold models and smooth models
is meaningful only if we consider a blow-up of the orbifold singularities.
This  corresponds to the fact that some of the twisted states
acquire a non trivial vacuum expectation value (vev). Since, generically,
twisted states are charged under the gauge symmetry, the blow up
corresponds to a modified Higgs mechanism.

The requirement that supersymmetry is preserved in the blow up implies that
the ``allowed'' blow up directions are flat directions of the moduli
potential. In the following we investigate the properties of such
flat directions and show the details of the blow up in the $SO(32)$ 
and $E_8 \times E_8$ models.

We avoid  the complete study
of the supergravity description of the models,
that would clarify the precise matching of the models beyond
the spectrum point of view, and would spread new light also on the fate of
the massive gauge symmetries, but that 
goes well beyond the purposes of present
paper.

We also comment that, typically, many flat directions are present
in an orbifold model, and we expect this to be true also for
the corresponding smooth model.
In this sense, given each smooth model, we search in the moduli
field configuration of its orbifold companion, in order to find a
vacuum (among the others) where the matching is complete.
A clear improvement of this picture would be a complete map
between the moduli spaces of each model, but such a result is
unfortunately not available at the moment.

\subsection{D-flatness and blow up of K3 orbifold models}
In a $d=6$ $\mathcal N = 1$ model, the potential for the hyper
multiplets is completely determined by the gauge interactions.
It contains only terms arising from the integration of auxiliary
fields in the vector superfield, and in this sense it is called
``D-term potential'', even though from a $d=4$ perspective it
contains both the D-term of the $d=4$ vector and the F-term of the 
chiral field that combines with the $d=4$ vector to form the $d=6$
vector.
We assume that the kinetic function of the hyper multiplets
is canonically normalized at the orbifold point, we comment later
about perturbations of this assumption.

Given an hyper multiplet with label $i$ we can organize the four real 
fields in a complex doublet $\Phi_i$, then, given $\sigma^a$ the three 
Pauli matrices, and $t^\alpha$ the generators of the gauge group we
define
\bea
D^{a,\alpha}= \Phi_i^\dagger \sigma^a t^\alpha_{ij} \Phi_j
\eea
and the scalar potential in six dimensions is then 
\bea
V=\sum_{a,\alpha} D^{a,\alpha}D^{a,\alpha}.
\eea
In the presence of $U(1)$ sectors we have $t^\alpha_{ij}=\delta_{ij} q_i$,
with $q_i$ the $U(1)$ charge, and we just drop the $\alpha$ index.
From a $d=4$ perspective we can see the standard D-term, arising
from $D^{3,\alpha}$ due to  $\sigma^3={\rm Diag}(1,-1)$
and the fact that a hyper multiplet contains two
complex scalars with opposite charges. 
Moreover, there are extra terms involving $D^{1,\alpha}$ and
$D^{2,\alpha}$.
As stated above, the latter are, from a $d=4$ perspective, the F-term
of the chiral multiplet that mixes with the $d=4$ vector multiplet to
form the $d=6$ vector multiplet. 

The existence of flat directions depends on the field content of the
model.
We check now the minimal content necessary to build flat directions,
and the kind of gauge symmetry breaking that is produced when a
vev is switched on along these directions.
The study of the general case goes beyond our scope, since
the models we studied contain only very specific representations
of $U(N)$, $SO(2N)$ and exceptional gauge groups, and only
some of them are necessary to explain the blow-ups, namely
the fundamental representations of $U(N)$ and $SO(2N)$ groups and the
spinorial representation of $SO(2N)$.
After the description of each flat direction we show how it is possible
to switch it on in some specific example and accommodate the
matching of the related orbifold model with the smooth case.
We notice that the matching can be made perfect in each of
the models, nevertheless, we also notice that not always the
``minimal'' blow up possibility is enough to achieve such a result.

\subsection{D-flatness in the $U(1)$ case: accommodated matching
in the 3-6a, 3c and III-VIa, IIIb models}

In the presence of a $U(1)$ symmetry and charged non-Abelian singlets in the
spectrum, a flat direction can exist only in case there
are at least two such singlets, independently on the charge.
Indeed, in case only one singlet is present, with charge $c$ and 
scalars $\Phi=(\phi_x,\phi_y)$, contrarily to what one would expect,
there is no flat direction. The naive expectation arises from the fact
that the two scalars have opposite $U(1)$ charges and 
\mbox{$D^3 =c (|\phi_x|^2-|\phi_y|^2)$}.
Unfortunately, the other contributions are such that
\bea
V =c^2 (|\phi_x|^2+|\phi_y|^2)^2,
\eea
and there are no flat directions.

In the presence of two hyper multiplets $\Phi_1=(\phi_{1x},\phi_{1y})$,
$\Phi_2=(\phi_{2x},\phi_{2y})$ of charge $q_1$ and $q_2$ we have
instead
\begin{equation}
\label{flatness}
V=\left[q_1 (|\phi_{1x}|^2+|\phi_{1y}|^2)+
q_2 (|\phi_{2x}|^2+|\phi_{2y}|^2)\right]^2
-4 q_1 q_2 |\phi_{1x}\phi_{2y}-\phi_{2x}\phi_{1y}|^2.
\end{equation}

Given this, we can choose $q_1=1$, $q_2=-1$, and  the potential
is a sum of positive terms
\begin{equation}
V=
\left(|\phi_{1x}|^2+|\phi_{1y}|^2-|\phi_{2x}|^2-|\phi_{2y}|^2\right)^2
+4|\phi_{1x}\phi_{2y}-\phi_{2x}\phi_{1y}|^2.
\end{equation}
The flat directions are then given by the conditions
\begin{equation}
|\phi_{1x}|^2+|\phi_{1y}|^2=|\phi_{2x}|^2+|\phi_{2y}|^2,\,\,
\phi_{1x}\phi_{2y}=\phi_{2x}\phi_{1y}.
\end{equation}
The second condition completely fixes one complex field in terms of the
others. Assume for a moment that $\phi_{2y}$ is nonzero, then
\begin{equation}
\label{redphi}
\phi_{1x}=\frac{\phi_{2x}\phi_{1y}}{\phi_{2y}},
\end{equation}
and replacing this in the first condition we obtain
\begin{equation}
\frac{|\phi_{2x}|^2+|\phi_{2y}|^2}{|\phi_{2y}|^2} |\phi_{1y}|^2=
|\phi_{2x}|^2+|\phi_{2y}|^2.
\end{equation}
This implies that $\phi_{2x}$ and $\phi_{2y}$ can be taken in
the whole space $\mathbb C^2$. If they are chosen away from the origin
of  $\mathbb C^2$, there is an extra condition $|\phi_{1y}|=|\phi_{2y}|$, so that
$\phi_{1y}$ is defined up to its phase, and $\phi_{1x}$ is well
defined in equation~(\ref{redphi}). If instead $\phi_{2x}=\phi_{2y}=0$,
both $\phi_{1y}$ and $\phi_{1x}$ must also be zero, given
the original condition.
If only $\phi_{2y}$ is zero and $\phi_{2x} \neq 0$, still  $\phi_{1y}$ must be zero given the
original conditions.
The flat directions are then locally given by the complex plane
$\mathbb C_1$ times the space $\mathbb C_2\times S^1$, where the
``radius'' of $S^1$ depends on the value of the $\mathbb C_2$
``coordinate''.

If a vev is switched on along one of the flat directions, the $U(1)$
vector boson becomes massive.
The gauge symmetry breaking is given by the kinetic terms
for the fields $\phi_{ix}$, $\phi_{iy}$.
We can rearrange these fields into a vector with four entries, 
$\tilde \phi_I$. The kinetic term is given by
\bea
M_{I,J}( \tilde \phi_K)
(\mathcal D {\tilde \phi_I})^\dagger \mathcal D \tilde \phi_J.
\eea
As mentioned above, we assume that the matrix $M$ is positive
definite at the orbifold point.
We can argue that it will remain such in case a ``small'' vev is
switched on along the flat direction.\footnote{Strictly speaking,
the $\phi$ dependence of $M$ would also affect the form of the
$D$-term potential and the D-flatness issue. 
Indeed, the extra corrections would in general produce new terms in
the potential, of higher order in the fields $\phi$.
We assume that we are allowed to neglect these extra terms, since,
under the assumption that $\langle \phi_i\rangle<<1$, they are 
parametrically smaller than those taken into account here.
Indeed, we expect that these extra terms can at most modify the exact shape
of the flat direction, without removing it.}

A positive mass term is generated for the $U(1)$ gauge
vector boson $A$ via the term
\bea
M_{I,J}(\langle \tilde \phi_K\rangle) \langle q_I \tilde \phi_I^*\rangle
\langle q_J \tilde \phi_J \rangle A^2 .
\eea

The computation above shows that if non-Abelian singlets with $U(1)$ charge
are present among the twisted states, a blow up with the $U(1)$
gauge vector becoming massive is actually possible, and a single
combination of the singlets is ``eaten'' in the process, so that the
matching between the spectra is achieved (we remind that the
computation on the smooth side actually provides only the difference
between the number of hypers and the number of massless
vector multiplets, and thus if in some process a vector boson becomes
massive, a corresponding hyper multiplet must disappear from the 
spectrum).
This does not imply that no blow up is possible in the presence
of a single hyper multiplet per fixed point: two hyper multiplets
coming from different fixed points do produce flat directions, meaning 
that the independent blow up of a single fixed point is impossible, but
a mutual blow up of many points is allowed. 

Given this we can argue that a prefect matching between orbifold
and smooth realizations in the  3-6a, 3c and III-VIa, IIIb cases can be
achieved, simply by switching on vevs along the flat directions
given by the twisted non-Abelian singlets with $U(1)$ charges present 
in the spectrum.
In all the other models a similar $U(1)$ breaking is also present,
but accompanied by a rank preserving gauge symmetry breaking
of the form $SU(N)\rightarrow SU(N-1) \times U(1)$ or $SO(2N)\rightarrow SU(N) \times U(1)$,
as we see in the following.

\subsection{D-flatness in the $SU(N)$ case: accommodated
matching in the 2a, 2c, 3b, 3d, 3e, 4a', 4b, 4e', 4g-i and IIa, IIIc, IIId,
IVb  models}

In the presence of an $SU(N)$ group, there is a D-term potential
corresponding to each generator of the gauge group.
The condition of D-flatness is more complicated, but, for
our purposes, we are allowed to consider only the case
of fields in the fundamental representation ${\bf N}$, i.e.
arrays $\Phi$ with $N$ entries $\Phi_i$. We consider only
the case that only $\langle \Phi_1\rangle\neq 0$. In this way,
only one D-flatness condition must be taken into account, completely
equivalent to the one studied in the $U(1)$ case.
This implies immediately that only in the presence of at least two
fields in the ${\bf N}$ representation, possibly also coming from different
fixed points, a flat direction can be built.
We conclude that the $SU(N)$ group is broken to $SU(N-1)$ (nothing in 
the $SU(2)$ case), and the broken vector bosons become massive.
The fields in the $\N$ representations are decomposed into 
$({\bf N-1})\oplus (\1)$, and two of the $({\bf N-1})$'s plus a singlet
are ``eaten'' in the process, consistently with the matching of
the spectrum.

This mechanism can be implemented,
in the 2a, 2c, 3b, 3d, 3e, 4a', 4b,
4e', 4g-i and IIa, IIIc, IIId, IVb models, to match the gauge groups
and the massless spectrum.

\subsection{D-flatness in the $SO(2N)$ case: accommodated
matching in the 2b, 3d, 4c-f, 4i, 6b models}

In all the 2b, 3d, 4c-f, 4i, 6b models the exact matching
between the orbifold and the smooth realization requires a mechanism
that breaks, on the orbifold side, some $SO(2N)$ group
to its subgroup $SU(N)\times U(1)$. Such a breaking can be explained in all
the models by the same mechanism, namely an Higgs mechanism
for a twisted field in the spinorial representation of the $SO(2N)$
gauge group. We show in the following that such a mechanism can be
introduced along flat directions of the potential.

It is possible to check that, among the twisted states of the models mentioned above, there are 
always spinorial representations that decompose under the breaking 
$SO(2N)\rightarrow SU(N)\times U(1)$ as described in appendix~\ref{App:Decomp}.

It is crucial to notice that these spinors can have negative chirality 
{\it only} in case $N$ is odd. Thus, under the decomposition, a
singlet is always present in the spectrum.
A vev of such a singlet is of course responsible for the symmetry
breaking, but we have to prove that such a vev can be switched on along
a flat direction of the potential.
In other words, we have to check the D-flatness condition for each of
the gauge group generators.
On the other hand, provided that we switch on only singlets, even
though they arise from different spinors, we have a single condition,
since the trace of any generator over the vacuum will be the same with 
a different weight (that is  zero if the singlet is not mapped into 
itself by the generator).
Thus, we have only copies of the same flatness condition~(\ref{flatness}), 
studied in the $U(1)$ case in presence of
at least two singlets, and so D-flatness is ensured in case
the decomposition of the multiplets present in a model provides (at least)
two singlets.

\section{Conclusions}
\label{Sec:Conclusions}

In this article, we have fully determined all $SO(32)$ and $E_8 \times E_8$ 
heterotic orbifold spectra on $T^4/\Z_N$ for $N=2,3,4$,
i.e. for $N=2,3$ added the missing $U(1)$ charges to the models of~\cite{Aldazabal:1997wi} 
and computed the $N=4$ spectra; for $N=6$, we have given some examples.
On the smooth side of $U(1)$ embeddings in $K3$, we have displayed a 
systematic treatment of $E_8$ line bundle embeddings and specialized  
on $U(1)$ bundles in the $SO(32)$ cases of~\cite{Honecker:2006dt}.

Using the field theoretical anomaly eight-forms, we have been able to map 
non-Abelian gauge groups at the orbifold point to those of the smooth phase with 
just one line bundle.
Up to the fact that at $\Z_2$ singularities, an additional $SU(2)$ gauge factor 
can occur or the rank of some gauge factor is enhanced by one according to
$SU(N) \rightarrow SU(N+1)$, $SU(N) \rightarrow SO(2N)$ or $SO(2N) \rightarrow SO(2N+2)$ in 
the orbifold limit, we find agreement for all $\Z_2$ spectra, all but one $\Z_3$
models for both $SO(32)$ and $E_8 \times E_8$ breakings, and part of the $\Z_4$ models with just one 
line bundle embedded.
In section~\ref{Sec:Blowup}, we have shown that the seeming mismatches in the 
orbifold point and $K3$ non-Abelian gauge groups disappear if the singularities are blown up, and the 
massless spectra are identical.
We have argued that for the remaining $\Z_4$ and $\Z_6$ models, similar embeddings 
with more than one line bundle will appear, and in the blow-up procedure more than rank 1 of the gauge group
will be broken.
Such a conclusion is supported by the fact that only in the $\Z_2$
and $\Z_3$ cases the orbifold fixed points are all equivalent, and
a single line bundle, switched on ``democratically'' among them,
can be enough for the matching: in the $\Z_{4}$ ($\Z_6$) case
the orbifold contains two (three) different fixed point species, namely, there
are both $\Z_4$ and $\Z_2$ ($\Z_6$, $\Z_2$ and $\Z_3$) fixed points.

The role of the $U(1)$ charges in all models clearly differs between the 
orbifold point and the smooth geometry, and the absence of any $ 2 \times 6$
factorization of the anomaly polynomial  at the orbifold point suggests that
the Abelian gauge bosons there remain all massless whereas in our class of
smooth embeddings, invariably some $U(1)$ gauge factor acquires a mass.
This phenomenon is most easily seen in case of the standard embeddings 3-6a
in $SO(32)$, which at the orbifold point have the same net number of
non-Abelian representations but differ in the $U(1)$ charge assignments of the
twisted sectors.
All these models have the same smooth match according to our identification
rule~(\ref{Eq:Id_Shifts}), and the field theoretical analysis of the blowing-up
procedure reveals that the $U(1)$s acquire a mass as needed.
The same applies to the  $E_8 \times E_8$ standard embeddings III-VIa.

At our level of matching the non-Abelian part of the spectra, the knowledge of the second Chern characters of 
the line bundles is sufficient, and we have digressed only briefly on potential explicit bundle realizations.

The six-dimensional analysis presented here is a particularly well tractable set-up due to the uniqueness of 
$K3$ and the strong conditions on the spectrum from gravitational and non-Abelian gauge 
anomaly cancellation. 
It remains to be seen if similar results can be obtained in heterotic $T^6/\Z_N$ and $T^6/(\Z_N \times \Z_M)$ 
compactifications to four dimensions.
It will also be interesting to see if the heterotic non-Abelian orbifolds have analogous matchings to embeddings with higher rank $U(n)$ bundles.

Last but not least, for the $SO(32)$ heterotic orbifolds listed here, threshold
corrections analogous to the ones computed for $E_8 \times E_8$
in~\cite{Stieberger:1998yi} as ${\cal N}=2$ sectors in four dimensions might
be evaluated and the moduli dependence beyond the leading order
in the geometric regime on Calabi-Yau three-folds extracted.


 \vskip 1cm
 {\noindent  {\large \bf Acknowledgments}}
\vskip 0.5cm
We are grateful for the hospitality of the workshop
``New Directions beyond the Standard Model in Field and String Theory'' at the 
the Galileo Galilei Institute, Florence, where this work was initiated.\\  
G.H. thanks R.~Blumenhagen, F.~Gmeiner, S.~Stieberger and T.~Weigand for
discussions and K.~Wendland for a useful communication;
M.T. thanks S.~Groot~Nibbelink and A.~Hebecker for discussions.
\vskip 2cm

\appendix

\section{Decomposition of representations upon gauge symmetry breaking}
\label{App:Decomp}

In this appendix, we list the types of gauge symmetry breaking required in order to compare 
the orbifold spectra with the smooth $K3$ compactifications. The blow-ups of orbifold singularities trigger 
the gauge symmetry breaking needed. 
\bea
\begin{array}{lcl}
SU(N+M) &\rightarrow & SU(N) \times SU(M)\times U(1)  
\\\hline\hline
{\bf N+M} &\rightarrow& (\N,\1)_M+(\1,\M)_{-N} 
\\\hline
{\bf \frac{(N+M)(N+M-1)}{2}}  &\rightarrow&  (\N,\M)_{-N+M} 
+ \left({\bf \frac{N(N-1)}{2}},\1\right)_{2M} 
+\left(\1,{\bf \frac{M(M-1)}{2}}\right)_{-2N}
\\\hline
{\bf (N+M)^2-1}  &\rightarrow&
(\N,\ov{\M})_{N+M}+ (\ov{\N},\M)_{-N-M}\\
&& + (\1,\1)_0 +  ({\bf N^2-1},\1)_0 + (\1,{\bf M^2-1})_0 
\\\hline
{\bf \frac{(N+M)(N+M-1)(N+M-2)}{6}} &\rightarrow&
({\bf \frac{N(N-1)(N-2)}{6}},\1)_{3M} 
+(\1,{\bf \frac{M(M-1)(M-2)}{6}})_{-3N}\\
&&
+ ({\bf \frac{N(N-1)}{2}},\M)_{2M-N}
+(\N,{\bf \frac{M(M-1)}{2}})_{M-2N}
\end{array} 
\nonumber
\eea
The most frequent case $SU(N+1) \rightarrow SU(N) \times U(1)$ is obtained by setting $M=1$ in the above breaking pattern. 
\bea
\begin{array}{lcl}
SO(2M)  &\rightarrow & SU(M) \times U(1) 
\\\hline\hline
{\bf 2M}  &\rightarrow & {\bf M}_1 + \ov{\bf M}_{-1} 
\\\hline
\Adj^{SO(2M)}   &\rightarrow & \Adj^{SU(M)}_0 + \1_0 + [\Anti^{SU(M)}_2 + c.c.]
\\\hline
\2_{\pm}^{M-1}   &\rightarrow & \left\{\begin{array}{c}
\sum_{k=0}^{[M/2]} \left( \begin{array}{c} M \\ 2k
\end{array}\right)_{-M/2 + 2k}
\\
\sum_{k=0}^{[(M-1)/2]} \left( \begin{array}{c} M \\ 2k+1
\end{array}\right)_{-M/2 + 2k + 1}
\end{array}\right.
\end{array} 
\nonumber
\eea
In the decomposition of the spinor representations, we have used the notation
\bea&&
\left(\begin{array}{c} M \\ 0 \end{array}\right) = {\bf 1},
\quad
\left( \begin{array}{c} M \\ 1 \end{array}\right) = {\bf M}, 
\quad
\left( \begin{array}{c} M \\ 2 \end{array}\right) = \Anti^{SU(M)}, 
\ldots\nonumber\\&&\ldots,
\left( \begin{array}{c} M \\ M-2 \end{array}\right) = 
\ov{\Anti}^{SU(M)}, 
\quad
\left( \begin{array}{c} M \\ M-1 \end{array}\right) = \ov{\bf M},
\quad
\left( \begin{array}{c} M \\ M \end{array}\right) = {\bf 1},
\nonumber
\eea
for the antisymmetric tensors of order $2k$ and $2k+1$.

Furthermore, in order to compute the $U(1)$ embeddings inside an $E_8$ gauge factor, we need the following decompositions:
\bea
\begin{array}{lcl}
SO(2N) &\rightarrow& SO(2N-2) \times U(1)
\nonumber\\\hline\hline
({\bf 2N}) &\rightarrow& ({\bf 2N-2})_0 + [(\1)_2 + c.c.]
\nonumber\\\hline
({\bf \frac{2N(2N-1)}{2}}) &\rightarrow& ({\bf \frac{(2N-2)(2N-3)}{2}})_0 +  (\1)_0 
\\
&&+ [ ({\bf 2N-2})_2 +c.c.]
\nonumber\\\hline
\left( \2^{N-1}_{\pm } \right) &\rightarrow& \left(\2^{N-2}_{\pm}\right)_1 + \left(\2^{N-2}_{\mp}\right)_{-1}
\end{array}
\nonumber
\eea


\end{document}